\documentclass[preprint,preprintnumbers,amsmath,amssymb]{revtex4}

\usepackage{graphicx}
\usepackage{amsmath,amssymb,amsthm,amsxtra,overpic,bbm,bm,epsfig,ulem,array}
\usepackage{color}
\usepackage{multirow}
\usepackage{rotating}

\graphicspath{{figures/}}

\makeatletter

\newcommand{\Rmnum}[1]{\expandafter\@slowromancap\romannumeral #1@}
\makeatother

\begin{document}

\title{Modifications to the neutrino mixing \\ from the $\mu$-$\tau$ reflection symmetry}

\author{Zhen-hua Zhao} \email{zhzhao@itp.ac.cn}

\affiliation{Department of Physics, Liaoning Normal University, Dalian 116029, China}

\begin{abstract}
The $\mu$-$\tau$ reflection symmetry serves as a unique basis for
understanding the observed neutrino mixing as it can lead us to the interesting results
$\theta^{}_{23} = \pi/4$ and $\delta = -\pi/2$ which stand close to the current
experimental results. But a precise measurement for $\theta^{}_{23}$ and
$\delta$ will probably force us to modify the neutrino mixing
$U^{(0)}$ from such a symmetry. Here we perform a study for modifications
to $U^{(0)}$ in the forms of $U^{(1)\dagger} U^{(0)}$ and $U^{(0)} U^{(1)}$
where $U^{(1)}= R^{(1)}_{ij}$ (for $ij = 12, 23$ and 13) with $R^{(1)}_{ij}$ denoting a real orthogonal rotation on the $ij$ plane.
\end{abstract}

\maketitle

\section{Introduction}

Thanks to the discovery of neutrino oscillations, it has been established that neutrinos have tiny
but non-zero masses and mix among different flavors \cite{pdg}. How to understand the smallness of neutrino masses and the pattern of neutrino mixing poses an important question for
particle physics. As is well known, the small neutrino masses can be naturally generated
by means of the seesaw mechanism \cite{seesaw}. And the neutrino mixing arises as $U=U^\dagger_l U^{}_\nu$ \cite{pmns} with $U^{}_l$ and $U^{}_\nu$ respectively resulting from
diagonalization of the charged lepton mass matrix $M^{}_l$ and the neutrino mass matrix $M^{}_\nu$. In the standard parametrization, $U$ reads
\begin{eqnarray}
U  = P^{}_l
\left( \begin{matrix}
c^{}_{12} c^{}_{13} & s^{}_{12} c^{}_{13} & s^{}_{13} e^{-{\rm i} \delta} \cr
-s^{}_{12} c^{}_{23} - c^{}_{12} s^{}_{23} s^{}_{13} e^{{\rm i} \delta}
& c^{}_{12} c^{}_{23} - s^{}_{12} s^{}_{23} s^{}_{13} e^{{\rm i} \delta}  & s^{}_{23} c^{}_{13} \cr
s^{}_{12} s^{}_{23} - c^{}_{12} c^{}_{23} s^{}_{13} e^{{\rm i} \delta}
& -c^{}_{12} s^{}_{23} - s^{}_{12} c^{}_{23} s^{}_{13} e^{{\rm i} \delta} & c^{}_{23}c^{}_{13}
\end{matrix} \right) P^{}_\nu \;,
\label{1}
\end{eqnarray}
where we have used the standard notations $c^{}_{ij} = \cos{\theta^{}_{ij}}$ and $s^{}_{ij} = \sin{\theta^{}_{ij}}$ for the mixing angles $\theta^{}_{ij}$ (for $ij = 12, 13$ and $23$). There is additionally one Dirac CP phase $\delta$, two Majorana CP phases $P^{}_\nu = {\rm Diag}(e^{{\rm i} \rho}, e^{{\rm i} \sigma}, 1)$ and three unphysical phases $P^{}_l = {\rm Diag}(e^{{\rm i} \phi^{}_1}, e^{{\rm i} \phi^{}_2}, e^{{\rm i} \phi^{}_3})$.
Owing to the accumulation of experimental data, some of these mixing parameters have
been measured to a good degree of accuracy. A recent global-fit result gives \cite{global}
\begin{eqnarray}
\sin^2{\theta^{}_{13}} = 0.02166 \pm 0.00075  \;, \hspace{1cm}
\sin^2{\theta^{}_{23}} = 0.441 \pm 0.024 \;, \hspace{1cm}
\delta = -(0.55 \pm 0.31)\pi \;,
\label{2}
\end{eqnarray}
for the normal neutrino mass ordering (NMO) $m^{}_1 < m^{}_2 < m^{}_3$,
\begin{eqnarray}
\sin^2{\theta^{}_{13}} = 0.02179 \pm 0.00076  \;, \hspace{1cm}
\sin^2{\theta^{}_{23}} = 0.587 \pm 0.022 \;, \hspace{1cm}
\delta = -(0.46 \pm 0.24)\pi \;,
\label{3}
\end{eqnarray}
for the inverted neutrino mass ordering (IMO) $m^{}_3 < m^{}_1 < m^{}_2$, and
\begin{eqnarray}
\sin^2{\theta^{}_{12}} = 0.306 \pm 0.012 \;,
\label{4}
\end{eqnarray}
for either neutrino mass ordering. But information about the Majorana CP phases is still lacking.

It is interesting to note that $\theta^{}_{12}$, $\theta^{}_{23}$ and the best-fit value of $\delta$ are close to some special values
\begin{eqnarray}
\sin^2{\theta^{}_{12}} = \frac{1}{3}  \;, \hspace{1cm}
\sin^2{\theta^{}_{23}} = \frac{1}{2} \ \left({\rm i.e.}, \ \theta^{}_{23} = \frac{\pi}{4} \right)  \;, \hspace{1cm}
\delta= - \frac{\pi}{2} \;.
\label{5}
\end{eqnarray}
(Of course, we still need a precise measurement for $\delta$ to see if it is really close to $-\pi/2$.) In view of these intriguing relations, it is natural to speculate that some
flavor symmetry has played a crucial role in shaping the observed neutrino mixing \cite{review}. A famous example is the $A^{}_4$ flavor symmetry \cite{A4} which can naturally produce the ever-popular tri-bimaximal (TBM) mixing \cite{TB}
\begin{eqnarray}
U^{}_{\rm TBM} = \frac{1}{\sqrt{6}} \left( \begin{matrix}
2 & \sqrt{2} & 0 \cr
-1 & \sqrt{2} & -\sqrt{3} \cr
-1 & \sqrt{2} & \sqrt{3}
\end{matrix} \right) \hspace{0.3cm} \Longrightarrow \hspace{0.3cm} \left\{
\begin{array}{l}
\vspace{0.2cm} \sin^2{\theta^{}_{12}} = \displaystyle \frac{1}{3}  \;, \\
\vspace{0.2cm} \theta^{}_{23} = \displaystyle \frac{\pi}{4}  \;, \\
\theta^{}_{13}= 0 \;.
\end{array} \right.
\label{6}
\end{eqnarray}
However, the observed $\theta^{}_{13} \simeq 0.15$ \cite{dyb}
requires a significant breaking of the flavor symmetries
that provide $\theta^{}_{13} =0$. Fortunately,
the $\mu$-$\tau$ reflection symmetry \cite{MTR,review2}
emerges as a unique alternative:
In the basis of $M^{}_l$ being diagonal, $M^{}_\nu$ keeps invariant under the transformations
\begin{eqnarray}
\nu^{}_e \leftrightarrow \nu^c_e \;, \hspace{1cm} \nu^{}_\mu \leftrightarrow \nu^c_\tau \;,
\hspace{1cm} \nu^{}_\tau \leftrightarrow \nu^c_\mu \;,
\label{7}
\end{eqnarray}
and is characterized by
\begin{eqnarray}
M^{}_{e\mu} = M^*_{e\tau} \;, \hspace{1cm} M^{}_{\mu\mu} = M^*_{\tau\tau}  \;, \hspace{1cm}
M^{}_{ee} \ {\rm and} \ M^{}_{\mu\tau} \ {\rm being \ real}  \;,
\label{8}
\end{eqnarray}
with $M^{}_{\alpha \beta}$ (for $\alpha, \beta = e, \mu$ and $\tau$) being the $\alpha\beta$-element of $M^{}_\nu$. The resulting neutrino mixing matrix $U^{(0)}$ (an analogue of $U$) allows for a non-zero $\theta^{}_{13}$ and features \cite{GL}
\begin{eqnarray}
&& \phi^{(0)}_{1} = \frac{\pi}{2} \;, \hspace{1cm} \phi^{(0)}_{ 2} = - \phi^{(0)}_{ 3}  \;,
\hspace{1cm} \theta^{(0)}_{ 23} =  \frac{\pi}{4} \;, \nonumber \\
&& \delta^{(0)} = \pm \frac{\pi}{2} \;, \hspace{1cm} \rho^{(0)}, \sigma^{(0)} = 0 \ {\rm or} \ \frac{\pi}{2} \;.
\label{9}
\end{eqnarray}
Explicitly, it can be written as
\begin{eqnarray}
U^{(0)}  = \frac{1}{\sqrt 2} \left( \begin{matrix}
{\rm i} \sqrt{2} c^{(0)}_{12} c^{(0)}_{13} e^{{\rm i} \rho^{(0)}}  & {\rm i} \sqrt{2} s^{(0)}_{12} c^{(0)}_{13} e^{{\rm i} \sigma^{(0)}}  & \sqrt{2} \eta^{}_\delta s^{(0)}_{13} \cr
- \left(s^{(0)}_{12} + {\rm i} \eta^{}_\delta c^{(0)}_{12} s^{(0)}_{13}  \right) e^{{\rm i} (\rho^{(0)} + \phi^{(0)})} & \left( c^{(0)}_{12} - {\rm i} \eta^{}_\delta s^{(0)}_{12} s^{(0)}_{13} \right) e^{{\rm i} (\sigma^{(0)}+ \phi^{(0)})}  & c^{(0)}_{13}  e^{{\rm i} \phi^{(0)}} \cr  \left(s^{(0)}_{12} - {\rm i} \eta^{}_\delta c^{(0)}_{12} s^{(0)}_{13} \right)  e^{{\rm i} (\rho^{(0)}-\phi^{(0)})}
& -\left(c^{(0)}_{12} + {\rm i} \eta^{}_\delta s^{(0)}_{12} s^{(0)}_{13} \right)  e^{{\rm i} (\sigma^{(0)}-\phi^{(0)})}  & c^{(0)}_{13}  e^{-{\rm i} \phi^{(0)}}
\end{matrix} \right)   \;,
\label{10}
\end{eqnarray}
with $\eta^{}_\delta = \pm 1$ (for $\delta^{(0)} = \pm \pi/2$), $\phi^{(0)} \equiv \phi^{(0)}_{ 2} = - \phi^{(0)}_{ 3}$, $c^{(0)}_{ij} = \cos{\theta^{(0)}_{ij}}$ and $s^{(0)}_{ij} = \sin{\theta^{(0)}_{ij}}$.
Because of these interesting predictions, the $\mu$-$\tau$ reflection symmetry
has been attracting a lot of interest \cite{MTRs}. It should be noted that one can further impose restrictions on $\theta^{(0)}_{12}$ and $\theta^{(0)}_{13}$ by combining this symmetry with other symmetries. For instance, the symmetry responsible for the TM1 or TM2 mixing (a mixing with the first or second column fixed to the TBM form \cite{TM12}) will lead us to \cite{RX}
\begin{eqnarray}
\left|U^{(0)}_{e1}\right|^2 = c^{(0)2}_{12} c^{(0)2}_{13} =  \frac{2}{3}
\hspace{0.5cm} {\rm or} \hspace{0.5cm}
\left|U^{(0)}_{e2}\right|^2 = s^{(0)2}_{12} c^{(0)2}_{13} =  \frac{1}{3} \;,
\label{11}
\end{eqnarray}
with $U^{(0)}_{\alpha i}$ being the $\alpha i$-element of $U^{(0)}$.

Nevertheless, a precise measurement for $\theta^{}_{23}$ and $\delta$ will probably point towards breakings of the $\mu$-$\tau$ reflection symmetry \cite{breakings}. Here we study the breaking effects of this symmetry in a situation where the resulting modified neutrino mixing matrix is parameterized in the form
of $U^{(1)\dagger} U^{(0)}$ or  $U^{(0)} U^{(1)}$ with $U^{(1)}$ being a correction matrix as has been done to many other neutrino mixing matrices (e.g., the TBM one) in the literature \cite{corrections}.
For the sake of predictability, we confine ourselves to the following simple scenario: $U^{(1)} = R^{(1)}_{ij}$ only consists of a single real orthogonal rotation
\begin{eqnarray}
R^{(1)}_{23} =  \left( \begin{matrix}
1 &  &  \cr
& c^{(1)}_{23} & s^{(1)}_{23} \cr
& -s^{(1)}_{23} & c^{(1)}_{23}
\end{matrix} \right)  \;, \hspace{1cm}
R^{(1)}_{13}  =  \left( \begin{matrix}
c^{(1)}_{13} &  & s^{(1)}_{13} \cr
& 1 &  \cr
-s^{(1)}_{13} &  & c^{(1)}_{13}
\end{matrix} \right)  \;, \hspace{1cm}
R^{(1)}_{12} =  \left( \begin{matrix}
c^{(1)}_{12} &  s^{(1)}_{12} & \cr
-s^{(1)}_{12} &  c^{(1)}_{12} & \cr
& & 1
\end{matrix} \right) \;,
\label{12}
\end{eqnarray}
with $c^{(1)}_{ij} = \cos{\theta^{(1)}_{ij}}$ and $s^{(1)}_{ij} = \sin{\theta^{(1)}_{ij}}$. The rotation angle $\theta^{(1)}_{ij}$ is allowed to take arbitrary values in the range $[0, \pi/2]$.

Before going into the details, we give the formula for extracting mixing parameters of the standard parametrization from the modified neutrino mixing matrix $U = U^{(1)\dagger} U^{(0)}$ or $U^{(0)} U^{(1)}$. On the one hand, one can get the mixing angles via
\begin{eqnarray}
s^{2}_{13} & = & \left|U^{}_{e3}\right|^2  \;, \hspace{1cm}
s^{2}_{12}  =  \frac{1}{c^{2}_{13}} \left|U^{}_{e2}\right|^2 \ \left({\rm or} \ c^{2}_{12}  =  \frac{1}{c^{2}_{13}} \left|U^{}_{e1}\right|^2 \right) \;, \nonumber \\
s^2_{23} & = &  \frac{1}{c^{2}_{13}} \left|U^{}_{\mu 3}\right|^2 \ \left({\rm or} \ c^{2}_{23}  =  \frac{1}{c^{2}_{13}} \left|U^{}_{\tau 3}\right|^2 \right)  \;,
\label{13}
\end{eqnarray}
with $U^{}_{\alpha i}$ being the $\alpha i$-element of $U^{}$.
On the other hand, the Dirac CP phase $\delta$ can be derived by letting the $U^{}_{\alpha i} U^{*}_{\alpha j}U^{*}_{\beta i} U^{}_{\beta j}$ (for $\alpha \neq \beta$ and $i \neq j$) obtained from the modified neutrino mixing matrix equal that obtained in the standard parametrization. For example, a choice of $\alpha =e, \beta= \mu, i=1$ and $j=3$ yields
\begin{eqnarray}
e^{{\rm i}\delta}  & = & - \frac{1}{c^{}_{12} s^{}_{12} c^{}_{23} s^{}_{23} c^2_{13} s^{}_{13}} \left(U^{}_{e1} U^{*}_{e3} U^{*}_{\mu 1} U^{}_{\mu 3} + c^{2}_{12} s^{2}_{23} c^2_{13} s^{2}_{13}\right) \;.
\label{14}
\end{eqnarray}
Then the Majorana CP phases $\rho$ and $\sigma$ are obtained as
\begin{eqnarray}
e^{{\rm i}\rho} = \frac{1}{c^{}_{12}c^{}_{13}s^{}_{13}} U^{}_{e1} U^*_{e3} e^{-{\rm i}\delta} \;, \hspace{1cm}
e^{{\rm i}\sigma} = \frac{1}{s^{}_{12}c^{}_{13}s^{}_{13}} U^{}_{e2} U^*_{e3} e^{-{\rm i}\delta} \;.
\label{15}
\end{eqnarray}

The rest part of this paper is organized as follows: In section \Rmnum{2} and \Rmnum{3} we study modifications to $U^{(0)}$ in the forms of $U^{(1)\dagger} U^{(0)}$ and $U^{(0)} U^{(1)}$, respectively. For each form, the cases of $U^{(1)} = R^{(1)}_{23}, R^{(1)}_{13}$ and $R^{(1)}_{12}$ are investigated in some detail. Finally, the main results are summarized in section \Rmnum{4}.

\section{Modifications to $U^{(0)}$ in the form of $U^{(1)\dagger} U^{(0)}$}

When $U^{(0)}$ is multiplied by a correction matrix from the left side, the resulting $\theta^{}_{23}$, $\theta^{}_{13}$, $\theta^{}_{12}$, $\delta$, $\Delta \rho \equiv \rho - \rho^{(0)}$ and $\Delta \sigma \equiv \sigma - \sigma^{(0)}$ are independent of $\rho^{(0)}$ and $\sigma^{(0)}$ but dependent on the unknown parameter $\phi^{(0)}$. Therefore, in this section we give the results for $\Delta \rho$ and $\Delta \sigma$ instead of $\rho$ and $\sigma$ and thus do not bother to consider the values of $\rho^{(0)}$ and $\sigma^{(0)}$. And the possible values of the parameters are presented as functions of $\phi^{(0)}$.

\subsection{$U^{(1)} = R^{(1)}_{23}$}

In the case of $U = R^{(1)\dagger}_{23} U^{(0)}$, $\theta^{}_{12}$ and $\theta^{}_{13}$ receive no contributions from $\theta^{(1)}_{23}$
(i.e., $\theta^{}_{12}  =  \theta^{(0)}_{12}$ and $\theta^{}_{13}  =  \theta^{(0)}_{13}$) while $\theta^{}_{23}$ is given by
\begin{eqnarray}
s^2_{23} = \frac{1}{2} - c^{(1)}_{23} s^{(1)}_{23} \cos{2\phi^{(0)}} \;.
\label{16}
\end{eqnarray}
Apparently, $\theta^{}_{23}$ remains $\pi/4$ for $\phi^{(0)}= \pi/4$ or $3\pi/4$. Taking the best-fit values $s^{2}_{23} = 0.441$ and $0.587$ in the NMO and IMO cases as illustration, we show the possible value of $s^{(1)}_{23}$ as a function of $\phi^{(0)}$ in the left figure of Fig. 1. In the numerical calculations throughout the paper, $\eta^{}_{\delta}= -1$ (for $\delta^{(0)} = -\pi/2$) and the best-fit values of $\theta^{}_{12}$ and $\theta^{}_{13}$ are used as input.
For $s^2_{23}<1/2$, $\phi^{(0)}$ is restricted to the range $[0, \pi/4]$ or $[3\pi/4, \pi]$ and further constrained by the condition $\cos{2\phi^{(0)}} \ge 2(1/2 - s^2_{23})$.
For $s^2_{23}>1/2$, $\phi^{(0)}$ is restricted to the range $[\pi/4, 3\pi/4]$ and further constrained by the condition $\cos{2\phi^{(0)}} \le 2(1/2 - s^2_{23})$. In most of the parameter space, one has $s^{(1)}_{23} \simeq |s^2_{23} - 1/2|$ or $c^{(1)}_{23} \simeq |s^2_{23} -1/2|$. But when $\phi^{(0)}$ approaches the bound values determined by $\cos{2\phi^{(0)}} = 2(1/2 - s^2_{23})$ (i.e., $\phi^{(0)} \simeq \pi/4- (1/2 - s^2_{23})$ or $3\pi/4 + (1/2 - s^2_{23})$), $s^{(1)}_{23}$ will quickly get close to $1/\sqrt{2}$.

\begin{figure}
\centering
\begin{minipage}[t]{0.49\textwidth}
\includegraphics[width=3.05in]{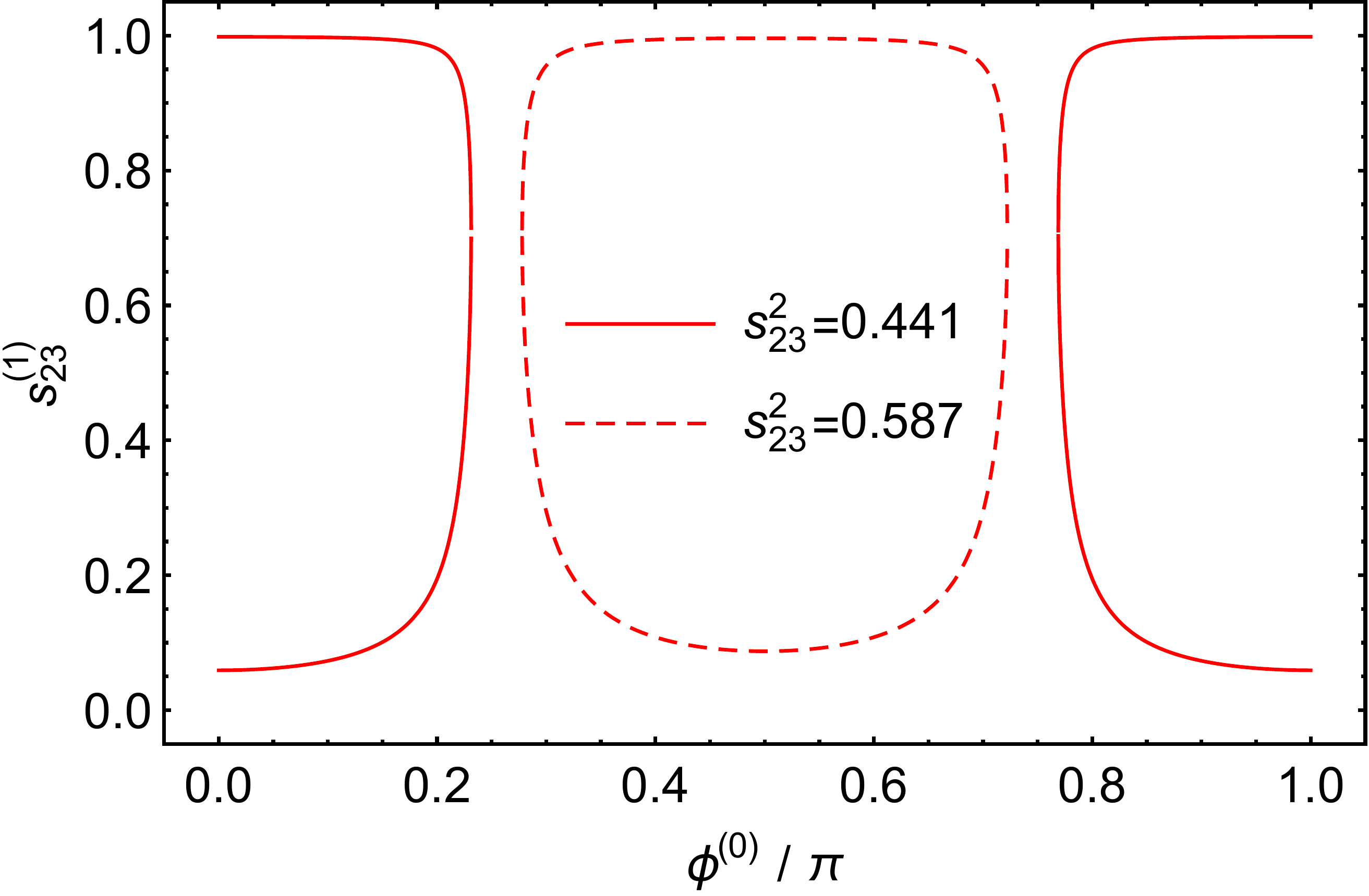}
\end{minipage}
\begin{minipage}[t]{0.49\textwidth}
\includegraphics[width=3.1in]{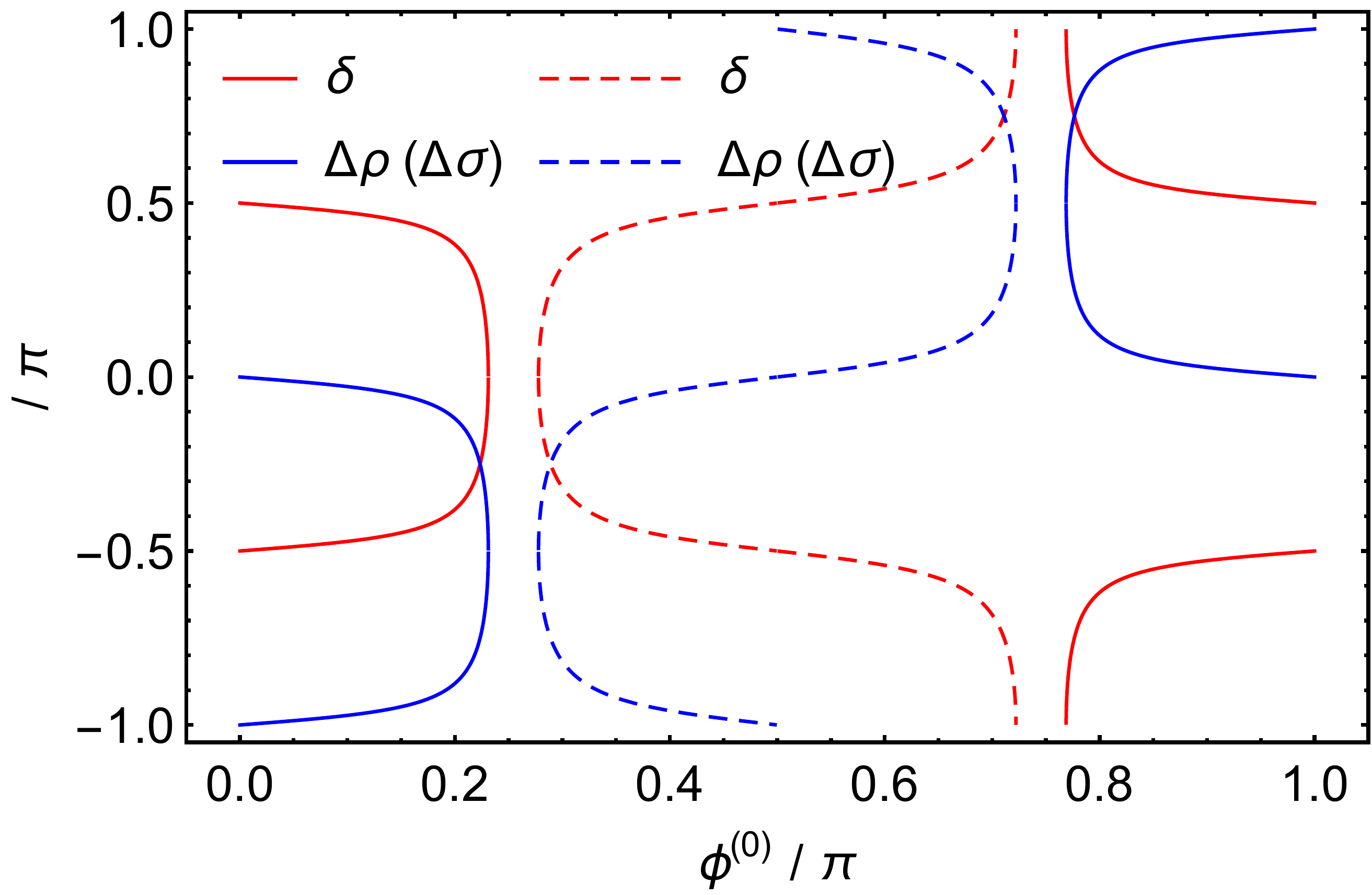}
\end{minipage}
\caption{ The possible values of $s^{(1)}_{23}$, $\delta$, $\Delta \rho$ and $\Delta \sigma$ as functions of $\phi^{(0)}$ for $s^{2}_{23} = 0.441$ (in full lines) and $0.587$ (in dashed lines) in the case of $U^{} = R^{(1)\dagger}_{23} U^{(0)}$. }
\end{figure}

One gets the Dirac CP phase as
\begin{eqnarray}
\sin{\delta} =  \eta^{}_\delta \frac{ c^{(1)2}_{23} - s^{(1)2}_{23} }{ 2 c^{}_{23} s^{}_{23} } \;, \hspace{1cm} \cos{\delta}
= -\eta^{}_\delta \frac{ c^{(1)}_{23} s^{(1)}_{23} }{  c^{}_{23} s^{}_{23}  } \sin{2\phi^{(0)}} \;.
\label{17}
\end{eqnarray}
In the right figure of Fig. 1, the possible value of $\delta$ is shown as a function of $\phi^{(0)}$ for $s^{2}_{23} = 0.441$ and $0.587$: $\delta$ lies close to $-\pi/2$ (for $s^{(1)}_{23} \simeq |s^2_{23} - 1/2|$) or  $\pi/2$ (for $c^{(1)}_{23} \simeq |s^2_{23} - 1/2|$) in most of the parameter space. But it will be around $0$ or $\pi$ when $\phi^{(0)}$ takes a value close to that fixed by $\cos{2\phi^{(0)}} = 2(1/2 - s^2_{23})$. Finally, one has $\Delta \rho = \Delta \sigma = -\delta +\eta^{}_\delta \pi/2$ for the Majorana CP phases.

\subsection{$U^{(1)} = R^{(1)}_{13}$}

In this case, the modified neutrino mixing matrix $U = R^{(1)\dagger}_{13} U^{(0)}$ leads us to the mixing angles as given by
\begin{eqnarray}
s^2_{23} & = & \frac{1}{2c^{2}_{13}} c^{(0)2}_{13} \;, \hspace{1cm}
s^{2}_{13}  =  c^{(1)2}_{13} s^{(0)2}_{13} - \sqrt{2} \eta^{}_\delta c^{(1)}_{13} s^{(1)}_{13} c^{(0)}_{13} s^{(0)}_{13} \cos{\phi^{(0)}} + \frac{1}{2} s^{(1)2}_{13} c^{(0)2}_{13} \;, \nonumber \\
s^2_{12} & = &  \frac{1}{c^{2}_{13}}
\left[\left(c^{(1)2}_{13} - s^2_{13} \right) s^{(0)2}_{12} -\sqrt{2} c^{(1)}_{13} s^{(1)}_{13} c^{(0)}_{12} s^{(0)}_{12} c^{(0)}_{13} \sin{\phi^{(0)}} + \frac{1}{2} s^{(1)2}_{13} \right] \;.
\label{18}
\end{eqnarray}
It is easy to see that $s^2_{23} \le 1/(2c^2_{13}) \simeq 0.51$, indicating that the global-fit result $s^{2}_{23} < 1/2$ in the NMO case is favored. The best-fit value $s^2_{23} = 0.441$ immediately gives $s^{(0)}_{13} \simeq 0.37$, while the possible values of $s^{(1)}_{13}$ and $s^{(0)}_{12}$ are shown as functions of $\phi^{(0)}$ in the left figure of Fig. 2. One finds that $\phi^{(0)}$ should fall in the range $[0.85 \pi, 1.15\pi]$ in order to give realistic mixing angles. (If $\eta^{}_\delta$ had been chosen as $+1$, then $\phi^{(0)}$ would take a value around 0.) $s^{(1)}_{13}$ and $s^{(0)}_{12}$ take values in the ranges $[0.31, 0.65]$ and $[0.31, 0.64]$, respectively. From these results one can see that in the case under consideration we need a large $\theta^{(1)}_{13}$ (much larger than the measured $\theta^{}_{13}$) to induce a sizable correction for $\theta^{}_{23}$. And a comparable $\theta^{(0)}_{13}$ (together with  $\phi^{(0)} \simeq \pi$) is needed to cancel the contribution of $\theta^{(1)}_{13}$ to $\theta^{}_{13}$ to an acceptable level.

The Dirac CP phase is obtained as
\begin{eqnarray}
&& 2c^{}_{12} s^{}_{12} c^{}_{23} s^{}_{23} s^{}_{13} \cos{\delta}
= -  \left(c^2_{12}- s^2_{12} \right) \left(s^2_{23} - \frac{1}{2} \right) -  s^2_{23} c^{2}_{13} \left(s^2_{12}  -  s^{(0)2}_{12} \right) \;, \nonumber \\
&& 2c^{}_{12} s^{}_{12} c^{}_{23} c^{}_{13} s^{}_{13} \sin{\delta}
= \sqrt{2} \eta^{}_{\delta} \left(c^{(1)2}_{13} - s^{(1)2}_{13}\right) c^{(0)}_{12} s^{(0)}_{12} c^{(0)}_{13} s^{(0)}_{13} - \eta^{}_{\delta} c^{(1)}_{13} s^{(1)}_{13} s^{(0)}_{13} \nonumber \\
&& \hspace{0.5cm}  \times \left(c^{(0)2}_{12} - s^{(0)2}_{12}\right) \sin{\phi^{(0)}}
+ c^{(1)}_{13} s^{(1)}_{13} c^{(0)}_{12} s^{(0)}_{12} \left(2 s^{(0)2}_{13} - c^{(0)2}_{13}\right) \cos{\phi^{(0)}} \;.
\label{19}
\end{eqnarray}
Its possible value is shown as a function of $\phi^{(0)}$ in the right figure of Fig. 2 for $s^2_{23} = 0.441$. The results show that the value of $\delta$ can saturate the range $[-\pi, \pi]$. Only in a small part of the parameter space (for $\phi^{(0)} \simeq 1.05 \pi$) can $\delta$ stay around $-\pi/2$. With the help of $\cos{\delta}$ and $\sin{\delta}$, $\cos{\Delta \rho}$ and $\cos{\Delta \sigma}$ are expressed as
\begin{eqnarray}
&& 2c^{}_{12} c^{}_{13} s^{}_{13} \cos{\Delta \rho}
=  \left[ s^{(1)2}_{13} s^{(0)}_{12} c^{(0)}_{13} -\sqrt{2} c^{(1)}_{13} s^{(1)}_{13} \left(\eta^{}_\delta s^{(0)}_{12} s^{(0)}_{13} \cos{\phi^{(0)}}-c^{(0)}_{12} \sin{\phi^{(0)}}\right) \right] \cos{\delta} \nonumber \\
&&  \hspace{0.5cm} + \left\{  \eta^{}_\delta \left(2c^{(1)2}_{13}-s^{(1)2}_{13}\right) c^{(0)}_{12} c^{(0)}_{13} s^{(0)}_{13} + \sqrt{2} c^{(1)}_{13} s^{(1)}_{13} \left[  \eta^{}_\delta s^{(0)}_{12} s^{(0)}_{13} \sin{\phi^{(0)}} - c^{(0)}_{12} \left(c^{(0)2}_{13}-s^{(0)2}_{13}\right) \right. \right. \nonumber \\
&& \hspace{0.5cm}\left. \left.  \times \cos{\phi^{(0)}}  \right]  \right\} \sin{\delta} \;, \nonumber \\
&& 2 s^{}_{12} c^{}_{13} s^{}_{13} \cos{\Delta \sigma}  =  \left[ -s^{(1)2}_{13} c^{(0)}_{12} c^{(0)}_{13} + \sqrt{2} c^{(1)}_{13} s^{(1)}_{13} \left(\eta^{}_\delta c^{(0)}_{12} s^{(0)}_{13} \cos{\phi^{(0)}}+s^{(0)}_{12} \sin{\phi^{(0)}}\right) \right] \cos{\delta} \nonumber \\
&&  \hspace{0.5cm} + \left\{  \eta^{}_\delta \left(2c^{(1)2}_{13}-s^{(1)2}_{13}\right) s^{(0)}_{12} c^{(0)}_{13} s^{(0)}_{13} - \sqrt{2} c^{(1)}_{13} s^{(1)}_{13} \left[  \eta^{}_\delta c^{(0)}_{12} s^{(0)}_{13} \sin{\phi^{(0)}} + s^{(0)}_{12} \left(c^{(0)2}_{13}-s^{(0)2}_{13}\right) \right. \right. \nonumber \\
&& \hspace{0.5cm} \left. \left. \times \cos{\phi^{(0)}} \right]   \right\} \sin{\delta} \;.
\label{20}
\end{eqnarray}
Here and in the following, $\sin{\Delta \rho}$ and $\sin{\Delta \sigma}$ (or $\sin{\rho}$ and $\sin{\sigma}$) can be read from $\cos{\Delta \rho}$ and  $\cos{\Delta \sigma}$ (or $\cos{\rho}$ and $\cos{\sigma}$) by making the replacements $\cos{\delta} \to -\sin{\delta}$ and $\sin{\delta} \to \cos{\delta}$. Compared to $\delta$, the Majorana CP phases are relatively stable against the correction effects: As shown by the right figure of Fig. 2, $\Delta \rho$ and $\Delta \sigma$ respectively vary in the ranges $[-0.29\pi, -0.11\pi]$ and $[0.03\pi, 0.11\pi]$.

When the $\mu$-$\tau$ reflection symmetry is combined with the symmetry responsible for the TM1 (TM2) mixing, there will be one more condition $c^{(0)2}_{12} c^{(0)2}_{13} = 2/3$ ($s^{(0)2}_{12} c^{(0)2}_{13} = 1/3$) that should be taken into account. In such a case, $s^{(0)}_{12}$ is fixed to $0.48$ ($0.62$). Consequently, one arrives at two solutions for $s^{(1)}_{13}$ and $\phi^{(0)}$
\begin{eqnarray}
&& s^{(1)}_{13} \simeq 0.34 \ (0.40) \;,  \hspace{1cm} \phi^{(0)} \simeq 1.09 \pi \ (0.87\pi) \;; \nonumber \\
&& s^{(1)}_{13} \simeq 0.64 \ (0.56)\;,  \hspace{1cm} \phi^{(0)} \simeq 0.96 \pi \ (0.87\pi) \;,
\label{21}
\end{eqnarray}
from Eq. (\ref{18}) and correspondingly two possible values for $\delta$, $\Delta \rho$ and $\Delta \sigma$
\begin{eqnarray}
&& \delta \simeq -0.55 \pi \ (-0.17 \pi) \;, \hspace{1cm} \Delta \rho \simeq -0.13\pi \ (-0.11 \pi) \;, \hspace{1cm} \Delta \sigma \simeq 0.04 \pi \ (0.08 \pi) \;; \nonumber \\
&& \delta \simeq 0.55 \pi \ (0.17 \pi) \;, \hspace{1cm} \Delta \rho \simeq -0.25 \pi \ (-0.17 \pi)\;, \hspace{1cm} \Delta \sigma \simeq 0.09 \pi \ (0.11 \pi)\;,
\label{22}
\end{eqnarray}
from Eqs. (\ref{19}, \ref{20}). The numbers outside (inside) the brackets are the results for the TM1 (TM2) case. Only in the TM1 case and when $s^{(1)}_{13}$ takes the smaller solution 0.34 can $\delta$ stay close to $-\pi/2$. Otherwise, it would be driven far away from $-\pi/2$. In comparison, the modifications of $\rho$ and $\sigma$ are relatively small.

\begin{figure}
\begin{minipage}[t]{0.49\textwidth}
\includegraphics[width=3.05in]{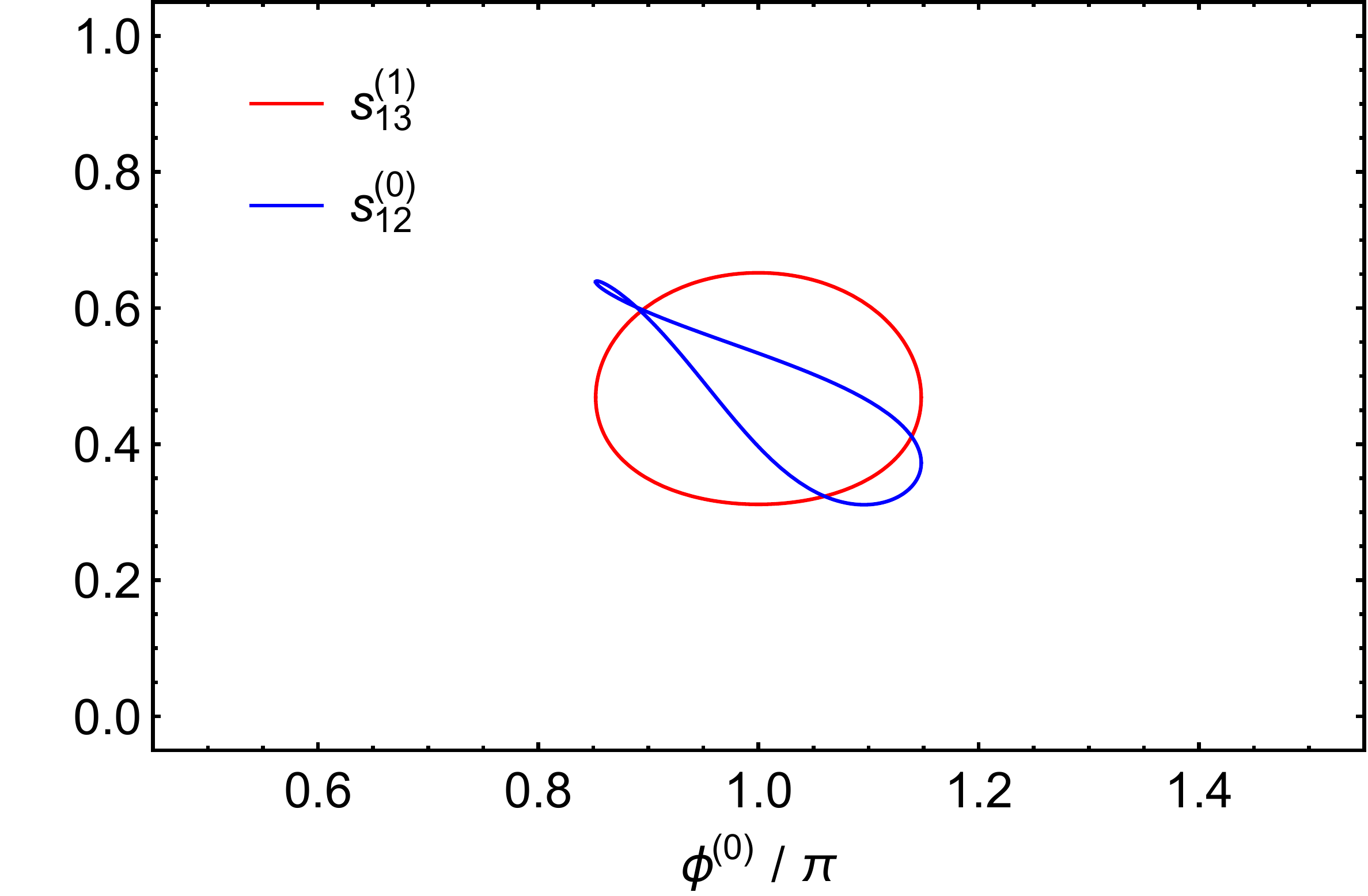}
\end{minipage}
\begin{minipage}[t]{0.49\textwidth}
\includegraphics[width=3.1in]{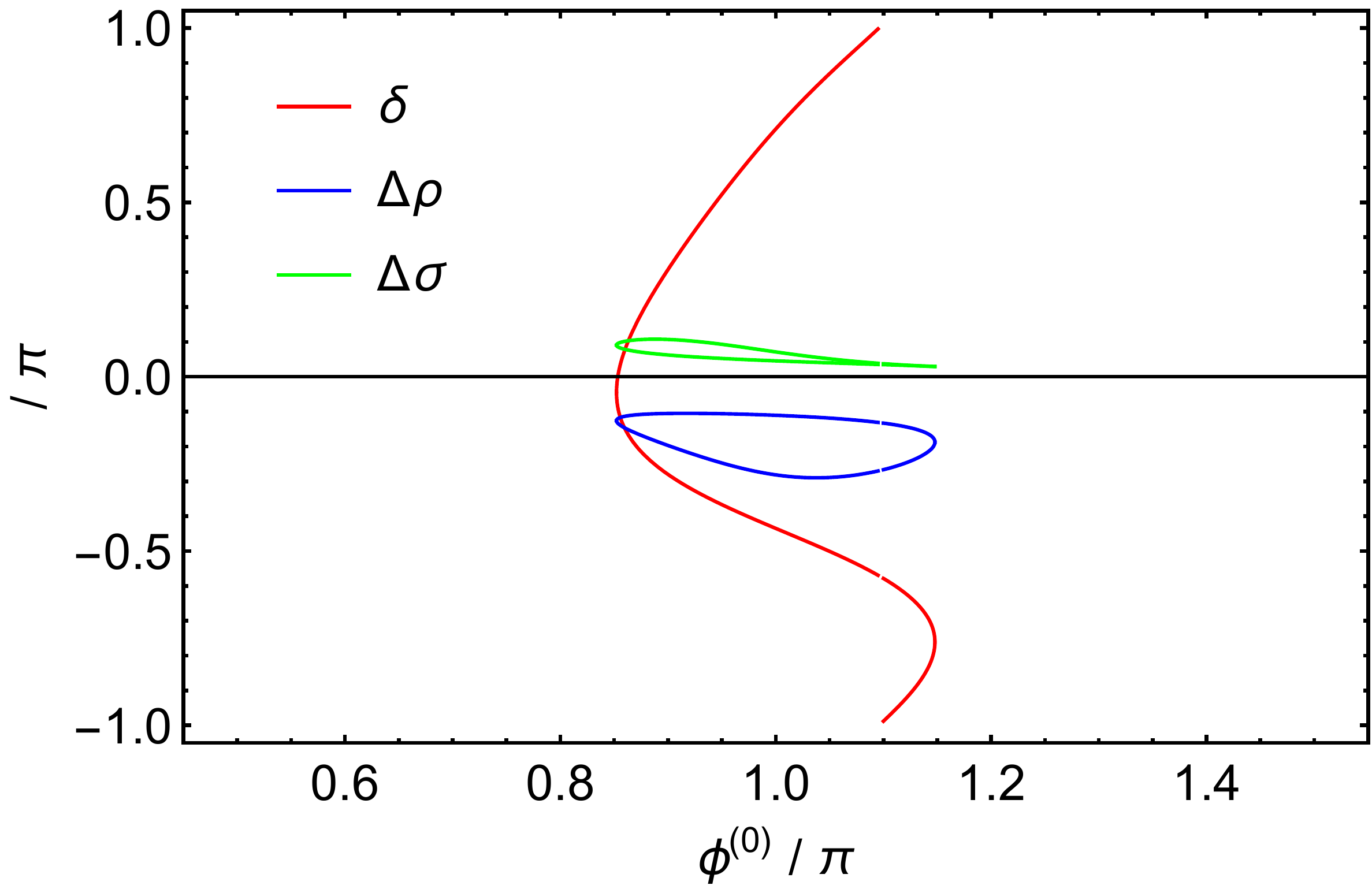}
\end{minipage}
\caption{The possible values of $s^{(1)}_{13}$, $s^{(0)}_{12}$, $\delta$, $\Delta \rho$ and $\Delta \sigma$ as functions of $\phi^{(0)}$ for $s^{2}_{23} = 0.441$ in the case of $U = R^{(1)\dagger}_{13} U^{(0)}$.}
\end{figure}

\subsection{$U^{(1)} = R^{(1)}_{12}$}

In this case, the modified neutrino mixing matrix $U = R^{(1)\dagger}_{12} U^{(0)}$ yields the mixing angles
\begin{eqnarray}
c^2_{23} & = &  \frac{1}{2c^{2}_{13} } c^{(0)2}_{13} \;, \hspace{1cm}
s^{2}_{13} =  c^{(1)2}_{12} s^{(0)2}_{13} - \sqrt{2} \eta^{}_\delta c^{(1)}_{12} s^{(1)}_{12} c^{(0)}_{13} s^{(0)}_{13} \cos{\phi^{(0)}} + \frac{1}{2} s^{(1)2}_{12} c^{(0)2}_{13} \;, \nonumber \\
s^2_{12} & =& \frac{1}{c^{2}_{13} } \left[ \left(c^{(1)2}_{12} -s^{2}_{13} \right) s^{(0)2}_{12}- \sqrt{2} c^{(1)}_{12} s^{(1)}_{12} c^{(0)}_{12} s^{(0)}_{12} c^{(0)}_{13}  \sin{\phi^{(0)}} + \frac{1}{2} s^{(1)2}_{12} \right]  \;.
\label{23}
\end{eqnarray}
This time we have $s^2_{23} \ge 1- 1/(2c^2_{13}) \simeq 0.49$, indicating that the global-fit result $s^{2}_{23} > 1/2$ in the IMO case is favored. The best-fit value $s^2_{23} = 0.587$ immediately gives $s^{(0)}_{13} \simeq 0.44$, while the possible values of $s^{(1)}_{12}$ and $s^{(0)}_{12}$ are shown as functions of $\phi^{(0)}$ in the left figure of Fig. 3. It turns out that $\phi^{(0)}$ should lie in the range $[0.87 \pi, 1.13\pi]$ so as to give realistic mixing angles. And $s^{(1)}_{12}$ and $s^{(0)}_{12}$ respectively lie in the ranges $[0.40, 0.71]$ and $[0.24, 0.62]$. So in this case one needs a large $\theta^{(1)}_{12}$ to induce a sizable correction for $\theta^{}_{23}$ and also a comparable $\theta^{(0)}_{13}$ (together with $\phi^{(0)} \simeq \pi$) to cancel its contribution to $\theta^{}_{13}$ by a large extent.

The Dirac CP phase is given by
\begin{eqnarray}
&& 2 c^{}_{12} s^{}_{12} c^{}_{23} s^{}_{23} s^{}_{13} \cos{\delta}
=  - \left(c^2_{12}- s^2_{12} \right) \left( s^2_{23} - \frac{1}{2} \right) +  c^2_{23} c^{2}_{13} \left(s^2_{12}  -  s^{(0)2}_{12} \right) \;, \nonumber \\
&& 2c^{}_{12} s^{}_{12} s^{}_{23} c^{}_{13} s^{}_{13} \sin{\delta}
= \sqrt{2} \eta^{}_{\delta} \left(c^{(1)2}_{12} - s^{(1)2}_{12}\right) c^{(0)}_{12} s^{(0)}_{12} c^{(0)}_{13} s^{(0)}_{13} - \eta^{}_{\delta} c^{(1)}_{12} s^{(1)}_{12} s^{(0)}_{13} \nonumber \\
&& \hspace{0.5cm}  \times \left(c^{(0)2}_{12} - s^{(0)2}_{12}\right) \sin{\phi^{(0)}}
+ c^{(1)}_{12} s^{(1)}_{12} c^{(0)}_{12} s^{(0)}_{12} \left(2 s^{(0)2}_{13} - c^{(0)2}_{13}\right) \cos{\phi^{(0)}} \;.
\label{24}
\end{eqnarray}
In the right figure of Fig. 3, we show its possible value as a function of $\phi^{(0)}$ for $s^2_{23} = 0.587$. As in the case of $U = R^{(1)\dagger}_{13} U^{(0)}$, the value of $\delta$ can saturate the range $[-\pi, \pi]$. And $\delta$ stays around $-\pi/2$ only
in a small part of the parameter space (for $\phi^{(0)} \simeq 1.06 \pi$). Given $\cos{\delta}$ and $\sin{\delta}$, $\cos{\Delta \rho}$ and $\cos{\Delta \sigma}$ appear as
\begin{eqnarray}
&& 2c^{}_{12} c^{}_{13} s^{}_{13} \cos{\Delta \rho}
=  \left[- s^{(1)2}_{12} s^{(0)}_{12} c^{(0)}_{13} + \sqrt{2} c^{(1)}_{12} s^{(1)}_{12} \left(\eta^{}_\delta s^{(0)}_{12} s^{(0)}_{13} \cos{\phi^{(0)}}-c^{(0)}_{12} \sin{\phi^{(0)}}\right) \right] \cos{\delta} \nonumber \\
&&  \hspace{0.5cm}  + \left\{ \eta^{}_\delta \left(2c^{(1)2}_{12}-s^{(1)2}_{12}\right) c^{(0)}_{12} c^{(0)}_{13} s^{(0)}_{13} + \sqrt{2} c^{(1)}_{12} s^{(1)}_{12} \left[  \eta^{}_\delta s^{(0)}_{12} s^{(0)}_{13} \sin{\phi^{(0)}} - c^{(0)}_{12} \left(c^{(0)2}_{13}-s^{(0)2}_{13}\right) \right. \right. \nonumber \\
&& \hspace{0.5cm}  \left. \left. \times \cos{\phi^{(0)}}  \right]  \right\} \sin{\delta} \;, \nonumber \\
&& 2 s^{}_{12} c^{}_{13} s^{}_{13} \cos{\Delta \sigma}  =  \left[ s^{(1)2}_{12} c^{(0)}_{12} c^{(0)}_{13} - \sqrt{2} c^{(1)}_{12} s^{(1)}_{12} \left(\eta^{}_\delta c^{(0)}_{12} s^{(0)}_{13} \cos{\phi^{(0)}}+s^{(0)}_{12} \sin{\phi^{(0)}}\right) \right] \cos{\delta} \nonumber \\
&&  \hspace{0.5cm} + \left\{  \eta^{}_\delta \left(2c^{(1)2}_{12}-s^{(1)2}_{12}\right) s^{(0)}_{12} c^{(0)}_{13} s^{(0)}_{13} - \sqrt{2} c^{(1)}_{12} s^{(1)}_{12} \left[ \eta^{}_\delta c^{(0)}_{12} s^{(0)}_{13} \sin{\phi^{(0)}} + s^{(0)}_{12} \left(c^{(0)2}_{13}-s^{(0)2}_{13}\right)  \right. \right. \nonumber \\
&& \hspace{0.5cm} \left. \left. \times \cos{\phi^{(0)}}  \right]   \right\} \sin{\delta} \;.
\label{25}
\end{eqnarray}
The Majorana CP phases are also relatively stable against the correction effects in the case under consideration: As shown by the right figure of Fig. 3, $\Delta \rho$ and $\Delta \sigma$ are  confined to the ranges $[0.14\pi, 0.35\pi]$ and $[-0.12\pi, -0.03\pi]$, respectively.

\begin{figure}
\begin{minipage}[t]{0.49\textwidth}
\includegraphics[width=3.05in]{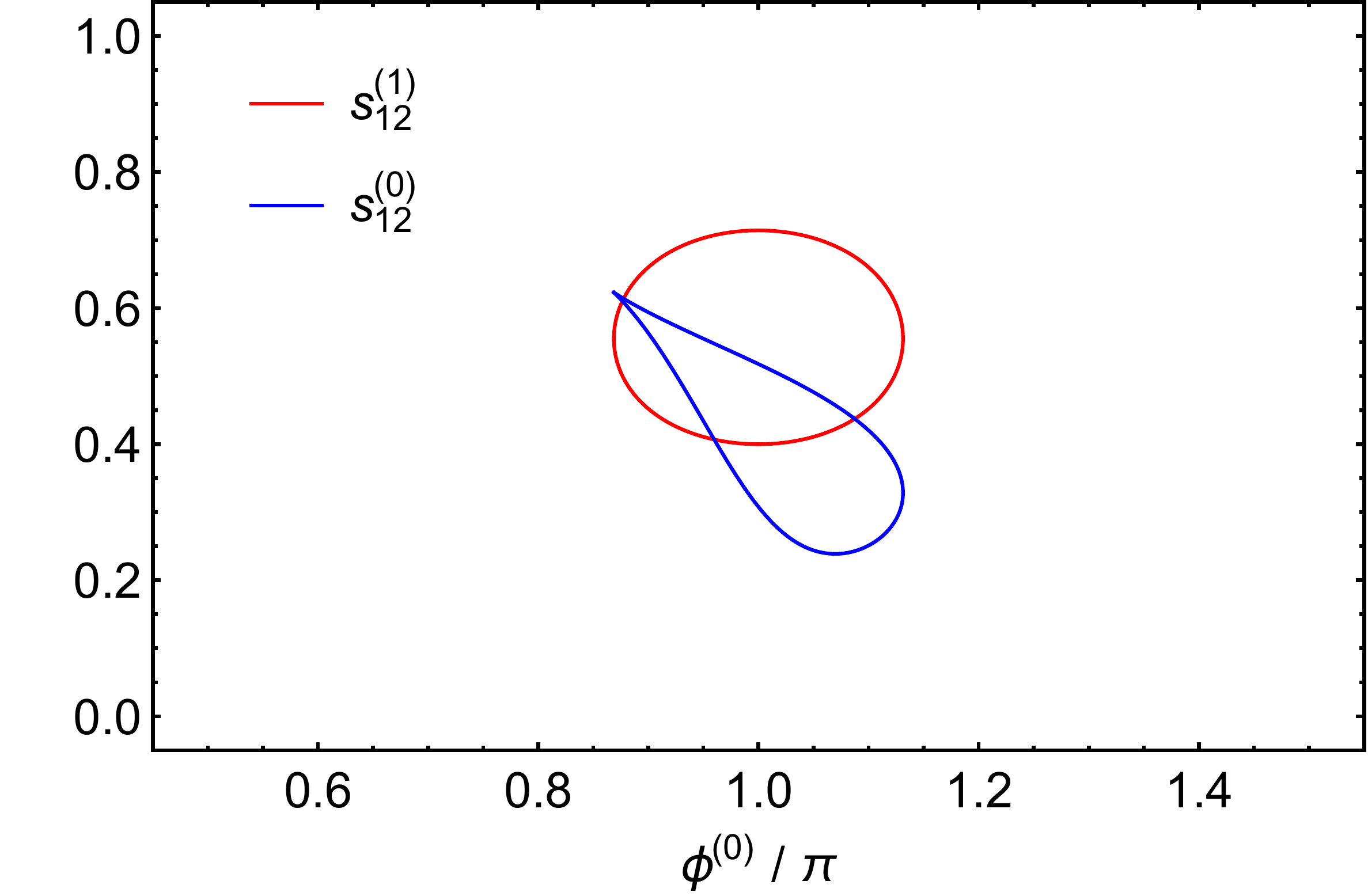}
\end{minipage}
\begin{minipage}[t]{0.49\textwidth}
\includegraphics[width=3.1in]{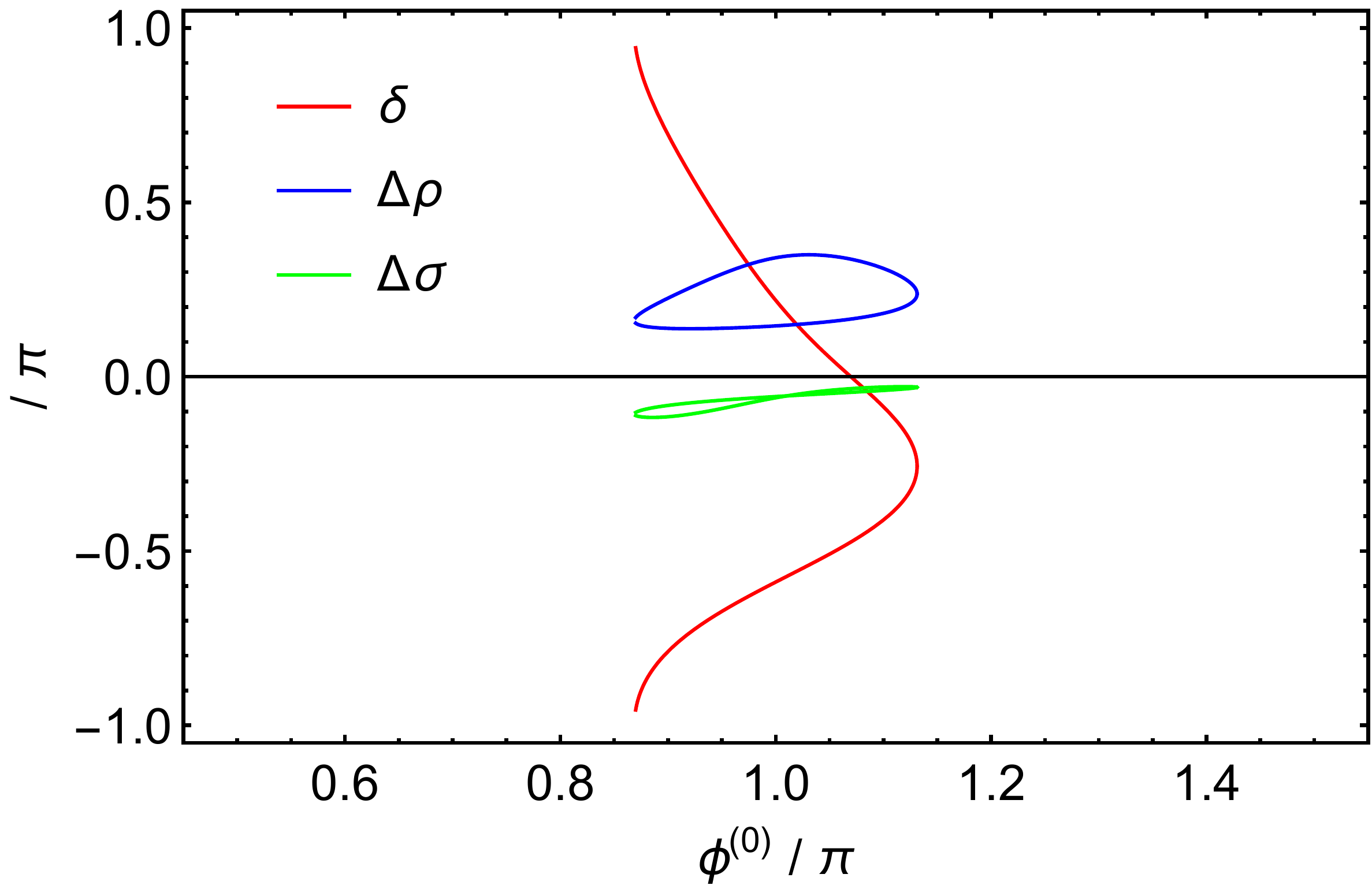}
\end{minipage}
\caption{The possible values of $s^{(1)}_{12}$, $s^{(0)}_{12}$, $\delta$, $\Delta \rho$ and $\Delta \sigma$ as functions of $\phi^{(0)}$ for $s^{2}_{23} = 0.587$ in the case of $U = R^{(1)\dagger}_{12} U^{(0)}$.}
\end{figure}

When the $\mu$-$\tau$ reflection symmetry is combined with the symmetry responsible for the TM1 (TM2) mixing, one will get $s^{(0)}_{12} \simeq 0.42$ ($0.63$), two solutions for $s^{(1)}_{12}$ and $\phi^{(0)}$
\begin{eqnarray}
&& s^{(1)}_{12} \simeq 0.45 \ (0.44) \;,  \hspace{1cm} \phi^{(0)} \simeq 1.10 \pi \ (0.86 \pi)\;; \nonumber \\
&& s^{(1)}_{12} \simeq 0.71 \ (0.56)\;, \hspace{1cm} \phi^{(0)} \simeq 0.96 \pi \ (0.87 \pi)\;,
\label{26}
\end{eqnarray}
from Eq. (\ref{23}) and correspondingly two possible values for $\delta$, $\Delta \rho$ and $\Delta \sigma$
\begin{eqnarray}
&& \delta \simeq -0.41 \pi \ (-0.87 \pi)\;, \hspace{1cm} \Delta \rho \simeq 0.18 \pi \ (0.12 \pi)\;, \hspace{1cm} \Delta \sigma \simeq -0.04 \pi \ (-0.08 \pi)\;; \nonumber \\
&& \delta \simeq 0.41 \pi \ (0.87 \pi)\;, \hspace{1cm} \Delta \rho \simeq 0.30 \pi \ (0.16 \pi) \;, \hspace{1cm} \Delta \sigma \simeq -0.09 \pi \ (-0.11 \pi)\;,
\label{27}
\end{eqnarray}
from Eqs. (\ref{24}, \ref{25}).
(The results for the TM2 case are obtained for $s^2_{23} = 0.565$, as the best-fit value $s^2_{23} = 0.587$ gives null results.) Similar to the results in the case of $U = R^{(1)\dagger}_{13} U^{(0)}$, $\delta$ can take a value around $-\pi/2$ only in the TM1 case and when $s^{(1)}_{12}$ takes the smaller solution. And the modifications of $\rho$ and $\sigma$ are relatively small.

\section{Modifications to $U^{(0)}$ in the form of $U^{(0)} U^{(1)}$}

When $U^{(0)}$ is multiplied by a correction matrix from the right side, the resulting mixing parameters will not depend on $\phi^{(0)}$ any more but depend on $\rho^{(0)}$ and $\sigma^{(0)}$. So in this section we have to deal with the values (0 or $\pi/2$) of $\rho^{(0)}$ and $\sigma^{(0)}$ case by case.

\subsection{$U^{(1)} = R^{(1)}_{23}$}

\begin{figure}
\begin{minipage}[t]{0.49\textwidth}
\includegraphics[width=3.1in]{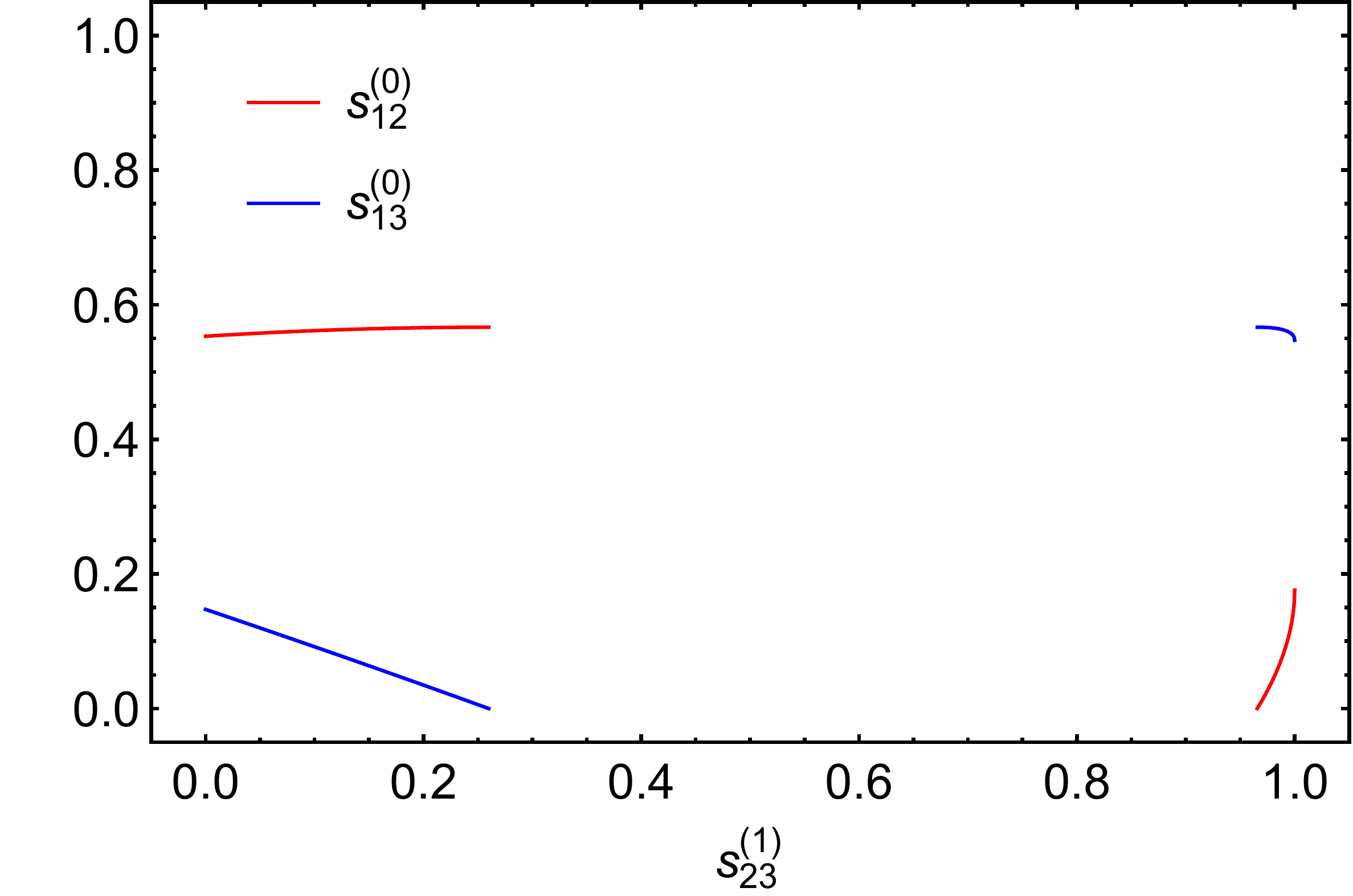}
\end{minipage}
\begin{minipage}[t]{0.49\textwidth}
\includegraphics[width=3.1in]{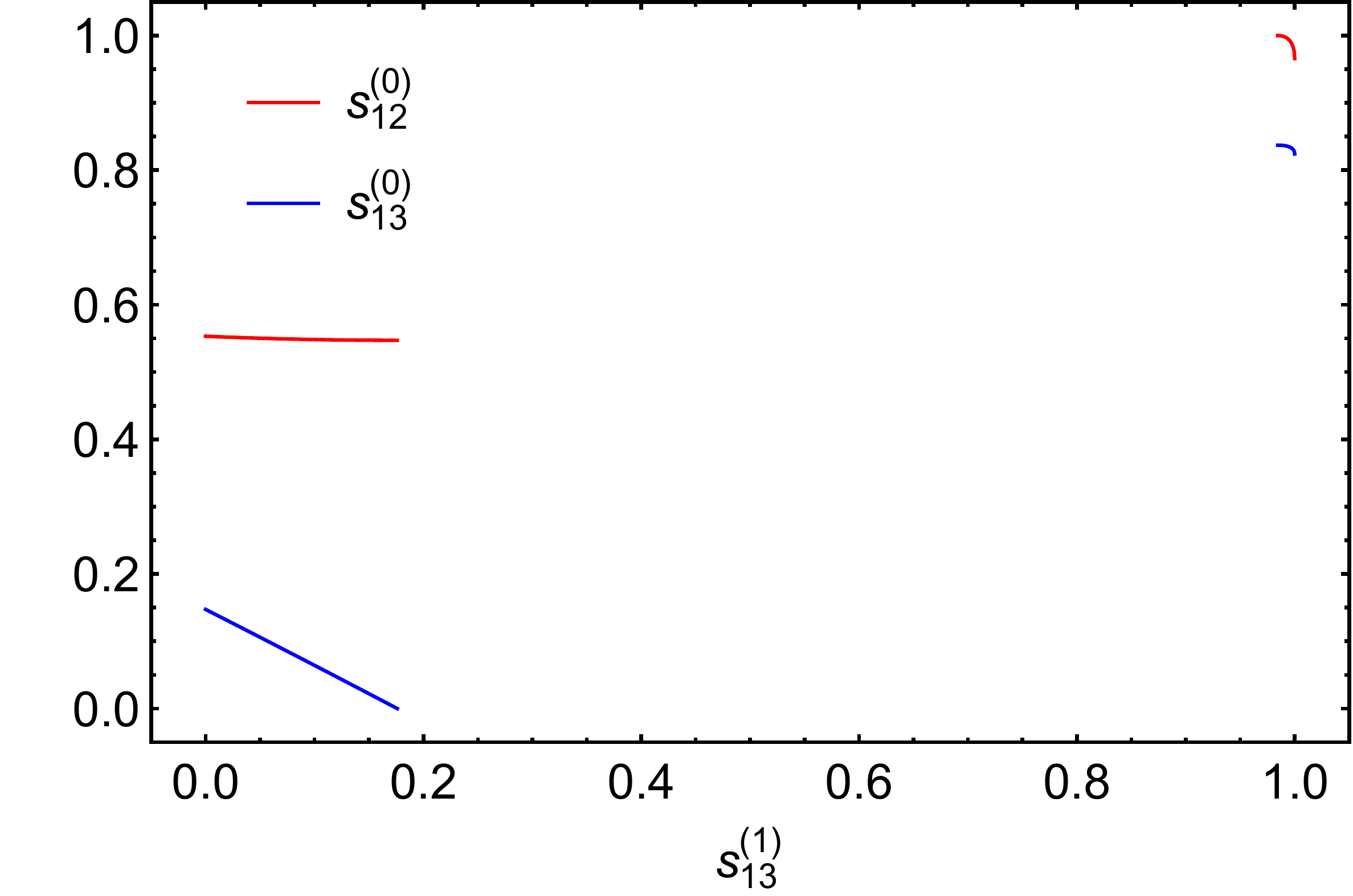}
\end{minipage}
\caption{The possible values of $s^{(0)}_{12}$ and $s^{(0)}_{13}$ as functions of $s^{(1)}_{23}$ (and $s^{(1)}_{13}$) for $\sigma^{(0)} =\pi/2$  (and $\rho^{(0)} =\pi/2$) in the case of $U = U^{(0)} R^{(1)}_{23}$ (and $U = U^{(0)} R^{(1)}_{13}$). }
\end{figure}

In the case of $U = U^{(0)} R^{(1)}_{23}$, the value of $\sigma^{(0)}$ is relevant for our study.
For $\sigma^{(0)} =0$, we obtain the mixing angles
\begin{eqnarray}
s^2_{23} & = & \frac{1}{2} + \frac{c^{}_{12}}{c^{}_{13}} c^{(1)}_{23} s^{(1)}_{23} \;, \hspace{1cm}
s^2_{13}  =   s^{(1)2}_{23} s^{(0)2}_{12} c^{(0)2}_{13} + c^{(1)2}_{23} s^{(0)2}_{13} \;, \hspace{1cm}
c^{2}_{12} =  \frac{1}{c^{2}_{13}} c^{(0)2}_{12} c^{(0)2}_{13}  \;,
\label{31}
\end{eqnarray}
the Dirac CP phase
\begin{eqnarray}
2 c^{}_{12}  s^{}_{12} c^{}_{23} s^{}_{23}  s^{}_{13}   \cos{\delta}
& = &  \left( s^2_{23}-\frac{1}{2} \right) \left( s^2_{12}-c^2_{12} s^2_{13} \right) \;, \nonumber \\
2s^{}_{12} c^{}_{23} s^{}_{23} c^{}_{13} s^{}_{13}  \sin{\delta} & = & \eta^{}_\delta s^{(0)}_{12} c^{(0)}_{13} s^{(0)}_{13} \left( c^{(1)2}_{23} - s^{(1)2}_{23} \right) \;,
\label{32}
\end{eqnarray}
and the Majorana CP phases
\begin{eqnarray}
s^{}_{13} \cos{\Delta \rho}
& = &  s^{(1)}_{23} s^{(0)}_{12} c^{(0)}_{13} \cos{\delta} + \eta^{}_\delta c^{(1)}_{23} s^{(0)}_{13} \sin{\delta} \;, \nonumber \\
s^{}_{12} c^{}_{13} s^{}_{13} \cos{\sigma}
& = &  c^{(1)}_{23} s^{(1)}_{23} \left(s^{(0)2}_{12} c^{(0)2}_{13}- s^{(0)2}_{13}\right) \cos{\delta} + \eta^{}_\delta s^{(0)}_{12} c^{(0)}_{13} s^{(0)}_{13} \sin{\delta} \;.
\label{33}
\end{eqnarray}
Taking the best-fit value $s^2_{23} = 0.587$ as input, one gets two solutions for $s^{(1)}_{23}$, $s^{(0)}_{12}$ and $s^{(0)}_{13}$
\begin{eqnarray}
&& s^{(1)}_{23} \simeq 0.10 \;, \hspace{1cm} s^{(0)}_{12} \simeq 0.56 \;, \hspace{1cm} s^{(0)}_{13} \simeq 0.14 \;; \nonumber \\
&& s^{(1)}_{23} \simeq 0.99 \;, \hspace{1cm} s^{(0)}_{12} \simeq 0.16 \;, \hspace{1cm} s^{(0)}_{13} \simeq 0.55 \;,
\label{34}
\end{eqnarray}
from Eq. (\ref{31}) and correspondingly two possible values for $\delta$, $\Delta \rho$ and $\sigma$
\begin{eqnarray}
&& \delta \simeq -0.38 \pi \;, \hspace{1cm} \Delta \rho \simeq 0.003 \pi \;, \hspace{1cm} \sigma \simeq -0.005 \pi \;; \nonumber \\
&& \delta \simeq 0.38 \pi \;, \hspace{1cm} \Delta \rho \simeq -0.50 \pi  \;, \hspace{1cm} \sigma \simeq -0.99 \pi \;,
\label{35}
\end{eqnarray}
from Eqs. (\ref{32}, \ref{33}). For $s^{(1)}_{23} \simeq 0.10$, the Dirac CP phase receives a relatively large correction while the modifications of other mixing parameters are much smaller. But for $s^{(1)}_{23} \simeq 0.99$ all the mixing parameters will suffer large modifications.

For $\sigma^{(0)} =\pi/2$, $\theta^{}_{23}$ remains $\pi/4$ while $\theta^{}_{13}$ and $\theta^{}_{12}$ are given by
\begin{eqnarray}
s^2_{13} = \left( \eta^{}_\delta c^{(1)}_{23} s^{(0)}_{13} - s^{(1)}_{23} s^{(0)}_{12} c^{(0)}_{13} \right)^2 \;, \hspace{1cm}
c^2_{12}  =  \frac{1}{c^2_{13}} c^{(0)2}_{12} c^{(0)2}_{13} \;.
\label{36}
\end{eqnarray}
We show the possible values of $s^{(0)}_{12}$ and $s^{(0)}_{13}$ as functions of $s^{(1)}_{23}$ in the left figure of Fig. 4: In the range $0 \le s^{(1)}_{23} \le 0.26$, $\theta^{(0)}_{12}$ approximates to the measured $\theta^{}_{12}$ while $\theta^{(0)}_{13}$ is small. In the range $0.96 \le s^{(1)}_{23} \le 1$, $\theta^{(0)}_{12}$ becomes small while $\theta^{(0)}_{13}$ gets close to the measured $\theta^{}_{12}$. As for the CP phases, there is $\sigma = \pi/2$ and
\begin{eqnarray}
\sin{\delta} & = & {\rm sgn} \left[ \left( \eta^{}_\delta c^{(1)}_{23} s^{(0)}_{13} - s^{(1)}_{23} s^{(0)}_{12} c^{(0)}_{13} \right) \left( \eta^{}_\delta s^{(1)}_{23} s^{(0)}_{13} + c^{(1)}_{23} s^{(0)}_{12} c^{(0)}_{13} \right) \right] \;, \nonumber \\
\cos{\Delta \rho} & = & {\rm sgn}\left( \eta^{}_\delta c^{(1)}_{23} s^{(0)}_{13} - s^{(1)}_{23} s^{(0)}_{12} c^{(0)}_{13} \right) \sin{\delta} \;.
\label{37}
\end{eqnarray}
In the range $0 \le s^{(1)}_{23} \le 0.26$, these results give $\delta = -\pi/2$ and $\Delta \rho = 0$. But in the range $0.96 \le s^{(1)}_{23} \le 1$ we are led to $\delta = \pi/2$ and $\Delta \rho = \pi$.

\subsection{$U^{(1)} = R^{(1)}_{13}$}

In the case of $U = U^{(0)} R^{(1)}_{13}$, the value of $\rho^{(0)}$ becomes relevant for our study.
For $\rho^{(0)} =0$, one acquires the mixing angles
\begin{eqnarray}
s^2_{23} & = & \frac{1}{2} - \frac{s^{}_{12}}{c^{}_{13}} c^{(1)}_{13} s^{(1)}_{13} \;, \hspace{1cm}
s^2_{13}  =  s^{(1)2}_{13} c^{(0)2}_{12} c^{(0)2}_{13} + c^{(1)2}_{13} s^{(0)2}_{13} \;, \hspace{1cm}
s^{2}_{12}  =  \frac{1}{c^{2}_{13} }s^{(0)2}_{12} c^{(0)2}_{13} \;,
\label{38}
\end{eqnarray}
the Dirac CP phase
\begin{eqnarray}
2 c^{}_{12} s^{}_{12} c^{}_{23} s^{}_{23}  s^{}_{13} \cos{\delta}
& = & \left(s^2_{23} -\frac{1}{2}\right) \left(s^{2}_{12} s^{2}_{13} -c^{2}_{12}\right) \;, \nonumber \\
2 c^{}_{12} c^{}_{23} s^{}_{23} c^{}_{13} s^{}_{13} \sin{\delta} & = &
\eta^{}_\delta c^{(0)}_{12} c^{(0)}_{13} s^{(0)}_{13} \left(c^{(1)2}_{13}  -s^{(1)2}_{13} \right) \;,
\label{39}
\end{eqnarray}
and the Majorana CP phases
\begin{eqnarray}
c^{}_{12} c^{}_{13} s^{}_{13} \cos{\rho}
& = &  c^{(1)}_{13} s^{(1)}_{13} \left(c^{(0)2}_{12} c^{(0)2}_{13}- s^{(0)2}_{13}\right) \cos{\delta} + \eta^{}_\delta c^{(0)}_{12} c^{(0)}_{13} s^{(0)}_{13} \sin{\delta} \;, \nonumber \\
s^{}_{13} \cos{\Delta \sigma}
& = &  s^{(1)}_{13} c^{(0)}_{12} c^{(0)}_{13} \cos{\delta} + \eta^{}_\delta c^{(1)}_{13} s^{(0)}_{13} \sin{\delta} \;.
\label{40}
\end{eqnarray}
Taking the best-fit value $s^2_{23} = 0.441$ as input, we get two solutions for $s^{(1)}_{13}$, $s^{(0)}_{12}$ and $s^{(0)}_{13}$
\begin{eqnarray}
&& s^{(1)}_{13} \simeq 0.11 \;, \hspace{1cm} s^{(0)}_{12} \simeq 0.55 \;, \hspace{1cm} s^{(0)}_{13} \simeq 0.12 \;; \nonumber \\
&& s^{(1)}_{13} \simeq 0.99 \;, \hspace{1cm} s^{(0)}_{12} \simeq 0.98 \;, \hspace{1cm} s^{(0)}_{13} \simeq 0.83 \;,
\label{41}
\end{eqnarray}
from Eq. (\ref{38}) and correspondingly two possible values for $\delta$, $\rho$ and $\Delta \sigma$
\begin{eqnarray}
&& \delta \simeq -0.30 \pi \;, \hspace{1cm}  \rho \simeq -0.007 \pi \;, \hspace{1cm} \Delta \sigma \simeq -0.002 \pi \;; \nonumber \\
&& \delta \simeq 0.30 \pi \;, \hspace{1cm}  \rho \simeq -0.99 \pi  \;, \hspace{1cm} \Delta \sigma \simeq -0.50 \pi \;,
\label{42}
\end{eqnarray}
from Eqs. (\ref{39}, \ref{40}). Similar to the results in the previous case,
for $s^{(1)}_{13} \simeq 0.11$ the Dirac CP phase receives a relatively large correction while the modifications of other mixing parameters are much smaller. And for $s^{(1)}_{23} = 0.99$ all the mixing parameters get remarkably modified.

For $\rho^{(0)} =\pi/2$, $\theta^{}_{23}$ remains $\pi/4$ while $\theta^{}_{13}$ and $\theta^{}_{12}$ turn out to be
\begin{eqnarray}
s^2_{13}  =  \left(\eta^{}_\delta c^{(1)}_{13} s^{(0)}_{13} - s^{(1)}_{13} c^{(0)}_{12} c^{(0)}_{13} \right)^2 \;, \hspace{1cm}
s^2_{12}  =  \frac{1}{c^2_{13}} s^{(0)2}_{12} c^{(0)2}_{13} \;.
\label{43}
\end{eqnarray}
The possible values of $s^{(0)}_{12}$ and $s^{(0)}_{13}$ are shown as functions of $s^{(1)}_{13}$ in the right figure of Fig. 4: In the range $0 \le s^{(1)}_{13} \le 0.18$, $\theta^{(0)}_{12}$ takes a value close to the measured one of $\theta^{}_{12}$ while $\theta^{(0)}_{13}$ is small. In the range $ 0.98 \le s^{(1)}_{13} \le 1$, $s^{(0)}_{12}$ and $s^{(0)}_{13}$ respectively become close to 1 and $c^{}_{12} \simeq 0.83$. As for the CP phases, there is $\rho = \pi/2$ and
\begin{eqnarray}
\sin{\delta} & = & {\rm sgn} \left[ \left( \eta^{}_\delta c^{(1)}_{13} s^{(0)}_{13} - s^{(1)}_{13} c^{(0)}_{12} c^{(0)}_{13} \right) \left( \eta^{}_\delta s^{(1)}_{13} s^{(0)}_{13} + c^{(1)}_{13} c^{(0)}_{12} c^{(0)}_{13} \right) \right] \;, \nonumber \\
\cos{\Delta \sigma} & = & {\rm sgn}\left( \eta^{}_\delta c^{(1)}_{13} s^{(0)}_{13} - s^{(1)}_{13} c^{(0)}_{12} c^{(0)}_{13} \right) \sin{\delta} \;.
\label{44}
\end{eqnarray}
One accordingly arrives at $\delta = -\pi/2$ and $\Delta \sigma = 0$ ($\delta = \pi/2$ and $\Delta \sigma = \pi$) in the range $0 \le s^{(1)}_{13} \le 0.18$ ($ 0.98 \le s^{(1)}_{13} \le 1$).

\subsection{$U^{(1)} = R^{(1)}_{12}$}

In the case of $U = U^{(0)} R^{(1)}_{12}$, both of the values of $\rho^{(0)}$ and $\sigma^{(0)}$ are relevant for our study. For $\rho^{(0)} = \sigma^{(0)} $, we simply have
\begin{eqnarray}
\theta^{}_{12} = \theta^{(0)}_{12} + \theta^{(1)}_{12}  \;,
\end{eqnarray}
with the other mixing parameters unchanged. For $\rho^{(0)} \neq \sigma^{(0)}$, $\theta^{}_{12}$ is given by
\begin{eqnarray}
s^2_{12} & = & s^{(1)2}_{12} c^{(0)2}_{12} + c^{(1)2}_{12} s^{(0)2}_{12} \;,
\end{eqnarray}
while $\theta^{}_{13}$ and $\theta^{}_{23}$ receive no corrections. The possible value of $s^{(0)}_{12}$ is shown as a function of $s^{(1)}_{12}$ in the left figure of Fig. 5 (and Fig. 6): $s^{(0)}_{12}$ decreases from $s^{}_{12} \simeq 0.55$ to 0 in the range $ 0 \le s^{(1)}_{12} \le 0.55$ and from 1 to $c^{}_{12} \simeq 0.83$ in the range $ 0.83 \le s^{(1)}_{12} \le 1$. On the other hand, for $[\rho^{(0)}, \sigma^{(0)}] = [0, \pi/2]$, one obtains the Dirac CP phase
\begin{eqnarray}
c^{}_{12} s^{}_{12} \cos{\delta} & = & \eta^{}_\delta c^{(1)}_{12} s^{(1)}_{12}  \;, \hspace{1cm}
c^{}_{12} s^{}_{12} \sin{\delta}  =  \eta^{}_\delta \left(c^{(1)2}_{12} - s^{(1)2}_{12} \right) c^{(0)}_{12} s^{(0)}_{12} \;,
\end{eqnarray}
and the Majorana CP phases
\begin{eqnarray}
c^{}_{12} \cos{\rho}
& = &  \eta^{}_\delta s^{(1)}_{12} s^{(0)}_{12} \cos{\delta} + \eta^{}_\delta c^{(1)}_{12} c^{(0)}_{12} \sin{\delta} \;, \nonumber \\
s^{}_{12} \cos{\sigma}
& = &  -\eta^{}_\delta c^{(1)}_{12} s^{(0)}_{12} \cos{\delta} + \eta^{}_\delta s^{(1)}_{12} c^{(0)}_{12} \sin{\delta} \;.
\end{eqnarray}
We show the possible values of them as functions of $s^{(1)}_{12}$ in the right figure of Fig. 5: In the range $ 0 \le s^{(1)}_{12} \le 0.55$ ($0.83 \le s^{(1)}_{12} \le 1$), $\delta$ decreases from $-\pi/2$ to $-\pi$ (from $\pi$ to $\pi/2$) while $\rho$ increases from 0 to $\pi/2$ (from 0 to $\pi/2$) and $\sigma$ takes a value around $\pi/2$ ($\pi$).
For $[\rho^{(0)}, \sigma^{(0)}] = [\pi/2, 0]$, the results become
\begin{eqnarray}
c^{}_{12} s^{}_{12} \cos{\delta} & = & -\eta^{}_\delta c^{(1)}_{12} s^{(1)}_{12} \;, \hspace{1cm}
c^{}_{12} s^{}_{12} \sin{\delta}  =  \eta^{}_\delta \left(c^{(1)2}_{12} - s^{(1)2}_{12}  \right) c^{(0)}_{12} s^{(0)}_{12} \;,
\end{eqnarray}
and
\begin{eqnarray}
c^{}_{12} \cos{\rho}
& = &  -\eta^{}_\delta c^{(1)}_{12} c^{(0)}_{12} \cos{\delta} - \eta^{}_\delta s^{(1)}_{12} s^{(0)}_{12} \sin{\delta} \;, \nonumber \\
s^{}_{12} \cos{\sigma}
& = &  -\eta^{}_\delta s^{(1)}_{12} c^{(0)}_{12} \cos{\delta} + \eta^{}_\delta c^{(1)}_{12} s^{(0)}_{12} \sin{\delta} \;.
\end{eqnarray}
As shown by the right figure of Fig. 6, in the range $ 0 \le s^{(1)}_{12} \le 0.55$ ($0.83 \le s^{(1)}_{12} \le 1$), $\delta$ increases from $-\pi/2$ to 0 (from 0 to $\pi/2$) while $\rho$ decreases from $\pi/2$ to 0 (from $\pi/2$ to 0) and $\sigma$ takes a value around 0 ($-\pi/2$).

\begin{figure}
\begin{minipage}[t]{0.49\textwidth}
\includegraphics[width=3.05in]{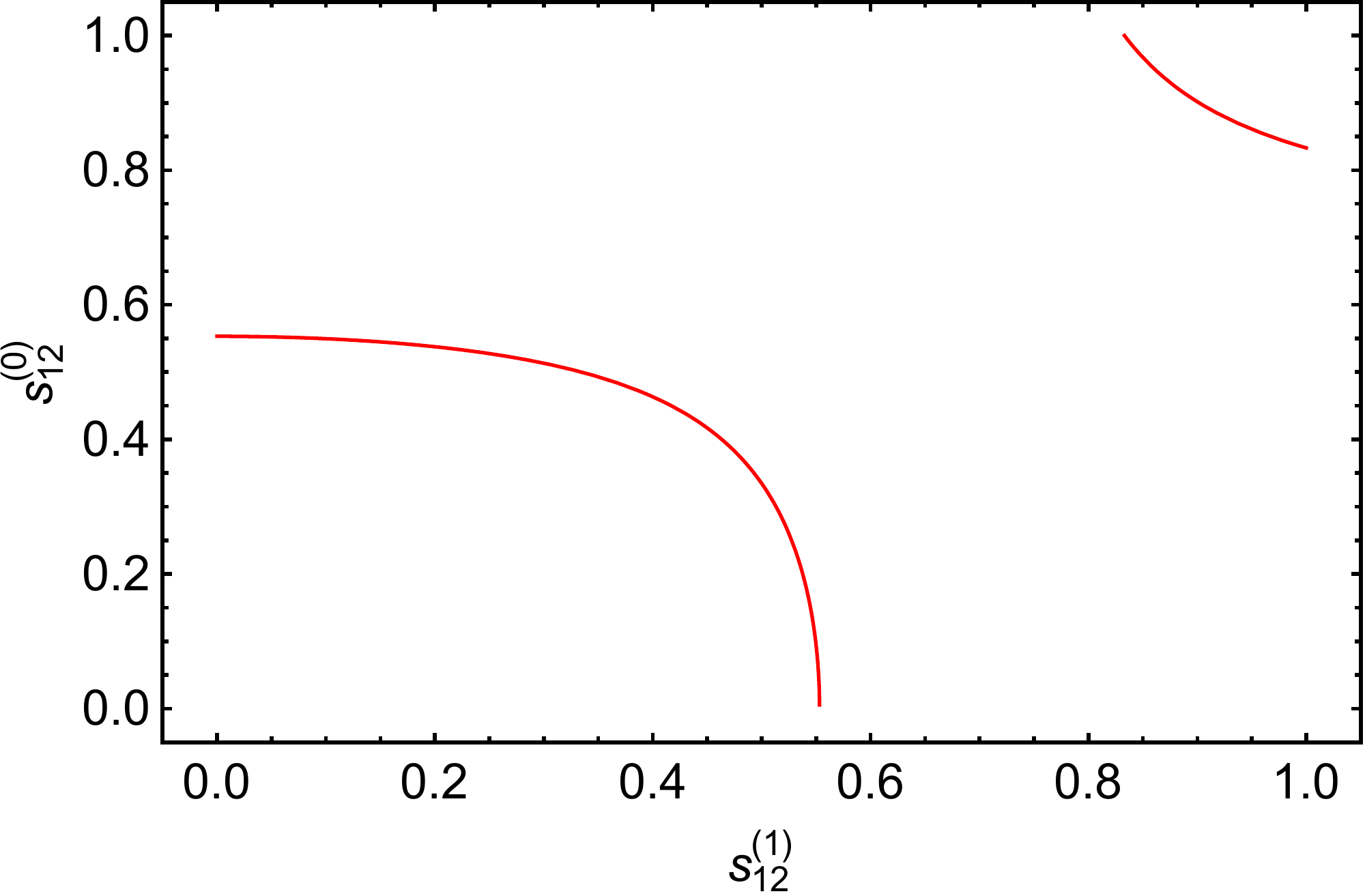}
\end{minipage}
\begin{minipage}[t]{0.49\textwidth}
\includegraphics[width=3.1in]{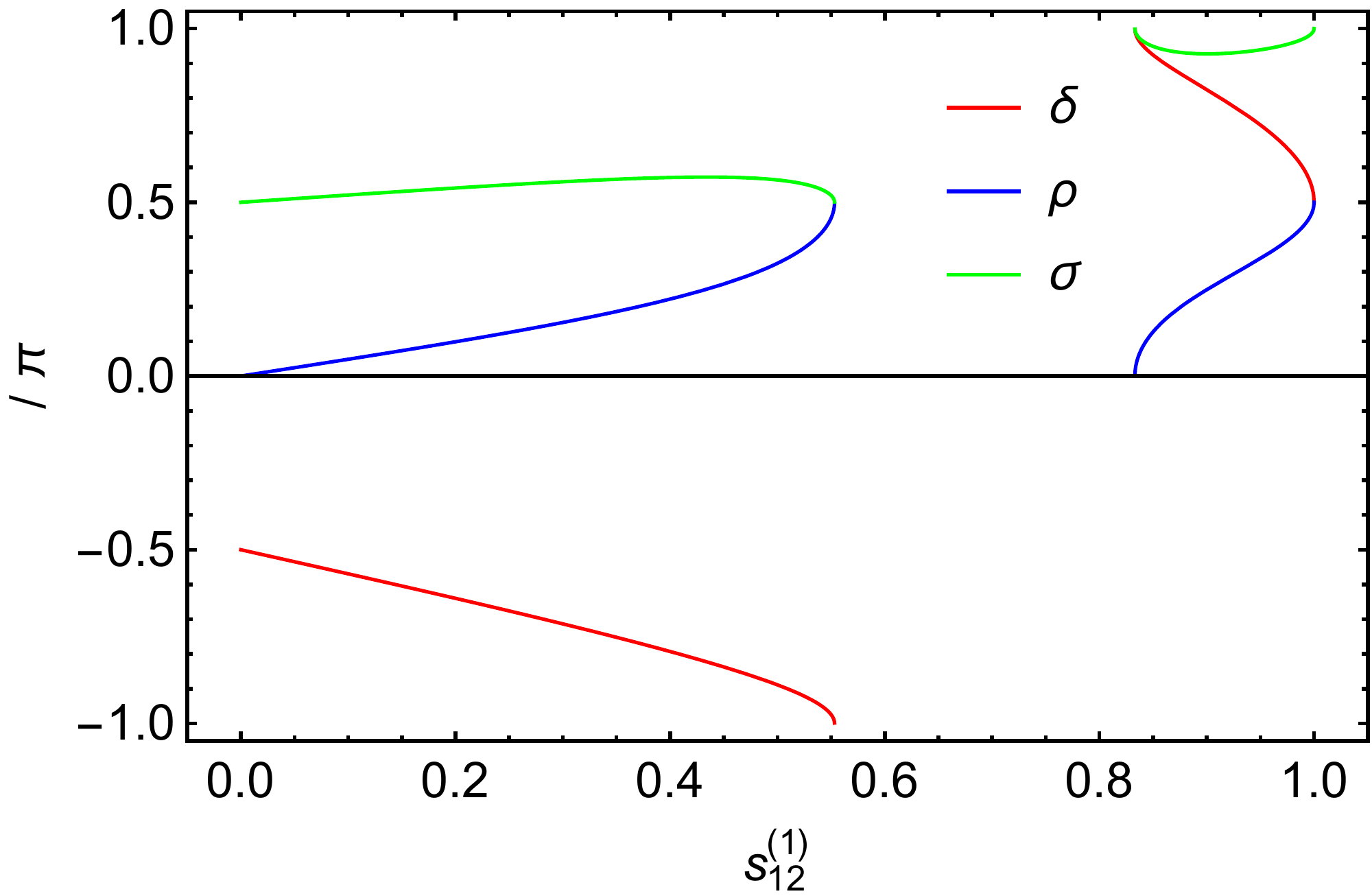}
\end{minipage}
\caption{The possible values of $s^{(0)}_{12}$, $\delta$, $\rho$ and $\sigma$ as functions of $s^{(1)}_{12}$ for $[\rho^{(0)}, \sigma^{(0)}] =[0, \pi/2]$ in the case of $U = U^{(0)} R^{(1)}_{12}$.}
\end{figure}

\begin{figure}
\begin{minipage}[t]{0.49\textwidth}
\includegraphics[width=3.05in]{s12r-angle-rho+sig-.pdf}
\end{minipage}
\begin{minipage}[t]{0.49\textwidth}
\includegraphics[width=3.1in]{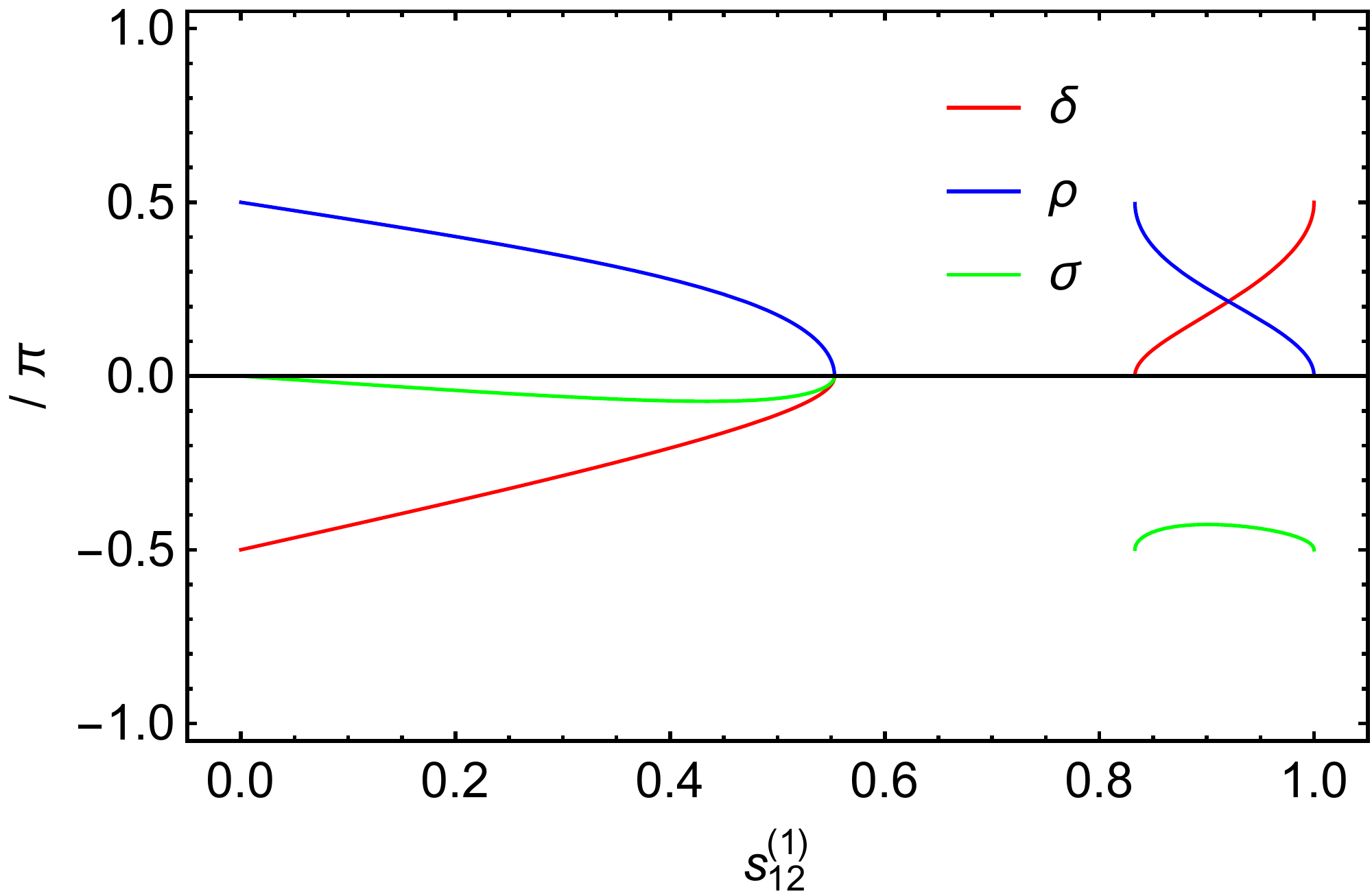}
\end{minipage}
\caption{The possible values of $s^{(0)}_{12}$, $\delta$, $\rho$ and $\sigma$ as functions of $s^{(1)}_{12}$ for $[\rho^{(0)}, \sigma^{(0)}] =[\pi/2, 0]$ in the case of $U = U^{(0)} R^{(1)}_{12}$.}
\end{figure}

\section{Summary and discussions}

In consideration of the interesting experimental results for the neutrino mixing parameters, explaining the observed neutrino mixing by using flavor symmetries is a worthwhile attempt. In this connection, the $\mu$-$\tau$ reflection symmetry provides a unique candidate as it can lead us to $\theta^{}_{23} = \pi/4$ and $\delta = -\pi/2$ which stand close to the current experimental results. But a precise measurement for $\theta^{}_{23}$ and $\delta$ will probably point towards breakings of this symmetry.
Hence we perform a study for modifications to the neutrino mixing $U^{(0)}$ given by this symmetry by multiplying it by a correction matrix $U^{(1)}$ from the left or right side. Here $U^{(1)}$ consists of a single rotation $R^{(1)}_{ij}$ where $\theta^{(1)}_{ij}$ can take arbitrary values in the range $[0, \pi/2]$. In addition, we also consider modifications to the neutrino mixing resulting from a combination of the $\mu$-$\tau$ reflection symmetry and the symmetry responsible for the TM1 (TM2) mixing.

For $U = U^{(1)\dagger} U^{(0)}$, the resulting $\theta^{}_{23}$, $\theta^{}_{13}$, $\theta^{}_{12}$, $\delta$, $\Delta \rho$ and $\Delta \sigma$ are independent of $\rho^{(0)}$ and $\sigma^{(0)}$ but dependent on the unknown parameter $\phi^{(0)}$. In the case of $U = R^{(1)\dagger}_{23} U^{(0)}$, among the mixing angles only $\theta^{}_{23}$ receives a correction from $\theta^{(1)}_{23}$. But it remains $\pi/4$ for $\phi^{(0)} = \pi/4$ or $3\pi/4$. For most of the parameter space of $\phi^{(0)}$, one has $s^{(1)}_{23} \simeq |s^2_{23} - 1/2|$ (or $c^{(1)}_{23} \simeq |s^2_{23} -1/2|$)
and correspondingly $\delta \simeq -\pi/2$ (or $\pi/2$). As for the Majorana CP phases, they are obtained as $\Delta \rho = \Delta \sigma = -\delta +\eta^{}_\delta \pi/2$. In the case of $U = R^{(1)\dagger}_{13} U^{(0)}$ ($U = R^{(1)\dagger}_{12} U^{(0)}$), the global-fit result $s^2_{23} < 1/2$ ($s^2_{23} > 1/2$) in the NMO (IMO) case is favored. And we need a large $\theta^{(1)}_{13}$ ($\theta^{(1)}_{12}$) to induce a sizable correction for $\theta^{}_{23}$ and a comparable $\theta^{(0)}_{13}$ (together with  $\phi^{(0)} \simeq \pi$) to cancel its contribution to $\theta^{}_{13}$ to an acceptable level. On the other hand, $\delta$ can saturate the range $[-\pi, \pi]$ and only stays around $-\pi/2$ for a small part of the parameter space of $\phi^{(0)}$, while the Majorana CP phases are relatively stable against the correction effects. When the $\mu$-$\tau$ reflection symmetry is combined with the symmetry responsible for the TM1 (TM2) mixing, the newly added condition will help us fix all the parameters. Only in the TM1 case and when $s^{(1)}_{13}$ ($s^{(1)}_{12}$) takes the smaller solution can $\delta$ keep close to $-\pi/2$.

For $U = U^{(0)} U^{(1)} $, the resulting mixing parameters are independent of $\phi^{(0)}$ but dependent on the values of $\rho^{(0)}$ and $\sigma^{(0)}$. In the case of $U = U^{(0)} R^{(1)}_{23}$ ($U = U^{(0)} R^{(1)}_{13}$), the value of $\sigma^{(0)}$ ($\rho^{(0)}$) is relevant. For $\sigma^{(0)} =0$ ($\rho^{(0)} = 0$), $s^{(1)}_{23}$ ($s^{(1)}_{13}$) should take a value of 0.1 or 0.99 in order to give the best-fit value of $s^2_{23}$. For $s^{(1)}_{23} \simeq 0.1$ ($s^{(1)}_{13} \simeq 0.1$), among the mixing parameters only $\delta$ acquires a relatively large correction. But for $s^{(1)}_{23} \simeq 0.99$ ($s^{(1)}_{13} \simeq 0.99$) all the mixing parameters get remarkably modified. For $\sigma^{(0)} =\pi/2$ ($\rho^{(0)} = \pi/2$), $\theta^{}_{23}$ remains $\pi/4$. In the case of $U = U^{(0)} R^{(1)}_{23}$, $\theta^{}_{23}$ does not receive any correction from $\theta^{(1)}_{12}$.

Finally, we discuss how the precise measurements of $\theta^{}_{23}$ and $\delta$ by the ongoing (e.g., T2K \cite{T2K} and NOvA \cite{NOvA}) and upcoming (e.g., DUNE \cite{DUNE} and T2HK \cite{T2HK}) neutrino oscillation experiments will impact the various cases considered in this study. Above all, we note that an up-to-date global-fit result \cite{global2} shows a preference for $\theta^{}_{23}$ in the upper octant. If this turns out to be the case, then the case of $U = R^{(1)\dagger}_{13} U^{(0)}$ which gives $s^2_{23} < 0.51$ will be ruled out. In the cases of $U = U^{(0)} R^{(1)}_{ij}$, the resulting mixing parameters are only dependent on one parameter (i.e., $\theta^{(1)}_{ij}$). One is thus left with a correlation between $\theta^{}_{23}$ and $\delta$ which can be verified or defied by the future precise measurements. But in the cases of $U = R^{(1)\dagger}_{ij} U^{(0)}$, the resulting mixing parameters are dependent on two parameters (i.e., $\theta^{(1)}_{ij}$ and $\phi^{(0)}$). This means that we have two free parameters to adjust to make the considered case consistent with the experimental results. So the precise measurements of $\theta^{}_{23}$ and $\delta$ are not necessarily competent to verify or defy these cases unless some information about the Majorana CP phases is provided from non-oscillatory processes (e.g., the neutrino-less double beta decays \cite{0nbb}).

\begin{acknowledgments}

This work is supported
in part by the National Natural Science Foundation of China under grant No. 11605081.

\end{acknowledgments}

\end{document}